\documentclass[3p,times,11pt]{elsarticle}
\usepackage{epsfig}
\usepackage{subfig}
\usepackage{graphicx}
\usepackage{amsmath}
\usepackage{amssymb}
\usepackage{lipsum}
\journal{Physica A}
\title{Quantum Dynamics of a Dissipative and Confined Cyclotron Motion}
\author[add1,add2]{Jishad Kumar\corref{cor1}}
\ead{kumar.jishad@gmail.com}
\address[add1]{Theoretical Condensed Matter Physics division, Saha Institute of Nuclear Physics, \\
1/AF Bidhan Nagar, Salt Lake City, Kolkata, West Bengal, India - 700064}
\address[add2]{Institute of Physics, Faculty of Mathematics and Physics, Charles University in Prague, \\
Ke Karlovu 5, 12116, Prague 2, Czech Republic}
\cortext[cor1]{Corresponding author email: kumar.jishad@gmail.com}
\cortext[cor2]{Present address}

\begin{document}

\date{Received: date / Revised version: date}

\begin{abstract}
We study the dissipative dynamics of a charged oscillator in a magnetic field by coupling (\textit{a la} Caldeira and Leggett) it to a heat bath consisting of non-interacting harmonic oscillators. We derive here the autocorrelation functions of the position and momentum and study its behavior at various limiting situations. The equilibrium (steady state) dispersions of position and momentum  are obtained from their respective autocorrelation functions. We analyse the equilibrium position and momentum dispersions at low and high temperatures for both low and high magnetic field strengths. We obtain the classical diffusive behavior (at long times) as well as the equilibrium momentum dispersion of the free quantum charged particle in a magnetic field, in the limit of vanishing oscillator potential $\omega_0$. We establish the relations between the reduced partition function and the equilibrium dispersions of the dissipative and confined cyclotron problem.
\end{abstract}
\begin{keyword}
Charged oscillator in a magnetic field, position and momentum dispersions, partition function
\end{keyword}

\maketitle
\section{Introduction}
\label{intro}
Quantum dissipation is ubiquitous in almost all fields of condensed matter physics\cite{weissbook,ingoldbook}. Dissipative quantum systems are well studied with the system-plus-reservoir (or heat bath) model, which is nowadays often referred as the Caldeira-Leggett (CL) model\cite{caldeira,leggett}. This model very well describes the underlying Brownian motion\cite{grabert,ingoldlect}. Because of the fact that the corresponding classical description is well understood, quantum Brownian motion has far reaching applications and is therefore still being investigated. The system-plus-bath model (or CL model) can be described as follows: a quantum subsystem of finite degree of freedom is coupled bilinearly to a reservoir (or heat bath) of non-interacting harmonic oscillators. Later, we integrate out the bath degrees of freedom to obtain a damped equation of motion for the system coordinates only. The infinite number of degrees of freedom in the heat bath allows for strong damping even if each individual bath oscillator couples only weakly to the system. This model has been discussed in the literature by many authors prior to Caldeira and Leggett, for harmonic systems\cite{maga,senit,fkm,ullersma} and anharmonic systems\cite{zwanzig}. 

Interesting effects occur when one study the dissipative effects in a magnetic field. For example, the discovery of non-classical transport of degenerate electron gas in the presence of strong disorder in quantized Hall effect\cite{prange}, temperature-dependent Hall effect in high-temperature superconductors\cite{ong} etc. In metals, the dissipative (or the damping) effect due to phase smearing reduces the amplitude of the de Haas - van Alphen oscillation (a non-linear oscillatory behavior of the magnetization of metals at very low temperatures and at very strong magnetic field strengths), and a dissipative phenomena involving carrier scattering produces the Shubnikov - de Haas effect (oscillations in the conductivity of metals at very low temperatures and very high magnetic fields) \cite{shoenberg}. With all these spectacular phenomena in the frontier, one may interest to choose a charged oscillator in a magnetic field  (actually a non-interacting electron gas confined in two dimensions) as the relevant quantum system of interest. This model of a dissipative charged oscillator in a magnetic field has been studied by many authors\cite{li1,li2,li3,hong,dattagupta,dattagupta2,malay3,jishad1,malaystat,jishad2,ban1,ban2}. In particular, this model was used to study the dissipative Landau   diamagnetism\cite{dattagupta,dattagupta2,jishad1} as well as to verify the validity of the third law of thermodynamics\cite{jishad1,jishad2,ban1}. 

In this paper, we study the dynamics of the damped charged harmonic oscillator in a magnetic field by analyzing the position and the velocity autocorrelation functions. Subsequently we derive the equilibrium dispersions of the position and momentum. In a previous paper \cite{jishad2} we have worked out the steady state equilibrium values of the position and momentum dispersions in order to compute the internal energy as the aim was to basically study the behavior of the specific heat at low and high temperatures under different limiting sequences within the framework of two different and distinct approaches to statistical mechanics, \textit{viz}., the \textit{Gibbsian} and the \textit{Einsteinian} approaches. Some of the details about the model and the basic mathematical framework we use here can be found in our previous paper\cite{jishad2}. But for the sake of continuity and completeness we describe here the model and the required mathematical details from our previous work. We study in detail the position autocorrelation function, its classical behavior at high temperatures and the quantum effects at zero temperature. Also we briefly sketch the fluctuation-dissipation relation in the framework of the dissipative Landau diamagnetism. It is pertinent to note here that, since we are computing the quantities at very long times compared to the system's characteristic time scale, the solution of the quantum Langevin equation is independent of initial transients so that the correlations are being calculated using the stationary solution. Hence we do not account here for the initial conditions and the contributions to the correlations from the initial values of the position and velocity. From the steady state results, we discuss here in detail about the temperature dependence of the position and momentum dispersions at low and high temperatures as well as its dependence on the bath spectrum at all temperatures. The dissipative charged particle dynamics (without a confinement potential) in a magnetic field has been studied in detail by Dattagupta and Singh \cite{singh}. The position auto-correlation function of the dissipative charged oscillator in a magnetic field as well as the classical diffusive behavior of the free quantum dissipative charged particle in a magnetic field was considered in detail using the retarded Green's function method by Li et al., \cite{li3}. We follow almost in a similar way and our results are in par with the results obtained by them. 

The total Hamiltonian representing the system-plus-bath model can be written as
\begin{equation}
\mathcal{H}=\mathcal{H}_S + \mathcal{H}_B + \mathcal{H}_{SB} ,
\label{full-ham}
\end{equation} 
where
\begin{equation}
\mathcal{H}_S = \frac{1}{2m}\bigg({\bf{p}}-\frac{e{\bf{A}}}{c} \bigg)^2 + \frac{1}{2}m\omega_0 ^2 {\bf{r}}^2 ,
\label{sys-ham}
\end{equation}
represents the system degrees of freedom,
\begin{equation}
\mathcal{H}_B = \sum_{j=1}^N \bigg\{\frac{{\bf{p}}_j ^2}{2m_j} + \frac{1}{2}m_j \omega_j ^2 {\bf{x}}_j ^2 \bigg\} ,
\label{bath-ham}
\end{equation}
represents the bath degrees of freedom and finally
\begin{equation}
\mathcal{H}_{SB} = -{\bf{r}}\sum_{j=1}^N C_j {\bf{x}}_j + {\bf{r}}^2 \sum_{j=1}^N \frac{C_j ^2}{2m_j \omega_j ^2} ,
\label{int-ham}
\end{equation}
is the Hamiltonian representing the system-bath interaction. The second term on the right hand side of Eq.(\ref{int-ham}) is included to suppress any unphysical renormalization of the system's potential due to the coupling with the bath, which thereby ensures the homogeneity of dissipation in all space and the translational invariance of the full Hamiltonian. Combining the three contributions we write the full many body Hamiltonian as
\begin{equation}
\label{hamiltonian}
\mathcal{H} = \frac{1}{2m}\bigg({\bf{p}} - \frac{e{\bf{A}}}{c} \bigg)^2 + \frac{1}{2} m\omega_0 ^2 {\bf{r}}^2 + \sum_{j=1} ^N \bigg\{ \frac{{\bf{p}}_j ^2}{2m_j} + \frac{1}{2} m_j \omega_j ^2 \bigg({\bf{x}}_j - \frac{C_j{\bf{r}}}{m_j\omega_j ^2}\bigg)^2 \bigg\} ,
\end{equation} 
where ${\bf{p}}$ and ${\bf{r}}$ are two-dimensional vectors representing the momentum and position coordinates of the system respectively. Similarly ${\bf{p}}_j$ and ${\bf{x}}_j$ are also two-dimensional vectors corresponding to the momentum and position coordinates of the bath oscillators respectively. The position and the momentum variables of the system and the bath are operators and they satisfy the following commutation relations
\begin{equation}
[{\bf{r}}_i , {\bf{p}}_k]=i\hbar\delta_{ik} ,~~[{\bf{x}}_{ji} , {\bf{p}}_{lk}] = i\hbar\delta_{jl} \delta_{ik} .
\label{commutators}
\end{equation}
${\bf{A}}$ is the magnetic vector potential. Here $m$ is the mass of the system and $m_j$ represents the masses of the heat bath oscillators. $\omega_0$ is the harmonic oscillator confinement frequency of the system and $\omega_j$s are the frequencies of the individual bath oscillators. Finally $C_j$ represents the coupling between the system and the bath. From Eq.(\ref{hamiltonian}) we obtain the generalized quantum Langevin equation for the system coordinate (operator) as \cite{zwanbook}:
\begin{equation}
\label{langevin1}
 m\ddot{{\bf{r}}}+\int_{0}^{t}dt^{\prime} \dot{{\bf{r}}}(t^\prime)
\gamma(t-t^\prime)-\frac{e}{c}(\dot{{\bf{r}}}\times{\bf{B}})+m\omega_0 ^2 {\bf{r}} = {\bf{F}}(t) .
\end{equation}
The integration in the forward direction of time breaks the time reversal invariance in the above equation which ensures the irreversibility in the problem. Note here that the Langevin equation is a gauge-independent one as it is devoid of the magnetic vector potential ${\bf{A}}$. Moreover the external magnetic field dependence in the equation is represented by the quantum version of the Lorentz force term. The memory friction function as well as the operator valued Gaussian random force are unchanged by the external magnetic field ${\bf{B}}$. The \textit{memory friction function} $\gamma(t)$, depends quadratically on the coupling parameter $C_j$ and  is given by
\begin{equation}
 \gamma(t) = \sum_{j} \frac{C^2 _{j}}{m_j \omega_j ^2}\cos(\omega_j t) .
\end{equation}
The noise ${\bf{F}}(t)$ depends explicitly on initial coordinates and the momenta of the bath oscillators, as is evident from the following equation
\begin{equation}
{\bf{F}}(t)=\sum_j \bigg\{C_j \bigg[{\bf{x}}_j (0)-\frac{C_j {\bf{r}}(0)}{m_j \omega_j
^2}\bigg]\cos(\omega_j t) + \frac{C_j {\bf{p}}_j (0)}{m_j \omega_j}\sin(\omega_j t)\bigg\} .
\label{friction}
\end{equation}
It is very much clear from Eq.(\ref{friction}) that the stochastic noise not only depends on the initial coordinates ${\bf{x}}_j (0)$ and momenta ${\bf{p}}_j (0)$ of the bath oscillators, but also on the initial condition of the system ${\bf{r}}(0)$ at time $t=0$. We can decompose the force operator into the following form
\begin{equation}
{\bf{F}}(t)=-m\gamma(t){\bf{r}}(0) + {\bf{K}}(t) ,
\end{equation}
where
\begin{equation}
{\bf{K}}(t)=\sum_j \bigg\{C_j {\bf{x}}_j (0) \cos(\omega_j t) + \frac{C_j {\bf{p}}_j (0)}{m_j \omega_j}\sin(\omega_j t)\bigg\} .
\end{equation}
This random force operator ${\bf{K}}(t)$ is a stationary Gaussian operator noise once the averages of the initial values ${\bf{x}}_j  (0)$ and ${\bf{p}}_j (0)$ are taken with respect to the initial equilibrium density matrix of the bath in canonical ensemble
\begin{equation}
\rho_B ^{eq} = \frac{1}{\mathcal{Z}}\exp\bigg\{-\beta \sum_{j=1} ^N
\bigg[\frac{{{\bf{p}}_j ^2} (0)}{2m_j} + \frac{1}{2}m_j \omega_j ^2 {\bf{x}}_j
  ^2 (0)  \bigg]  \bigg\},
\label{bath-density-matrix}
\end{equation}
where $\mathcal{Z}$ is the partition function of the bath. Because of the spurious initial transient slip term $\gamma(t){\bf{r}}(0)$ in Eq.(\ref{langevin1}), the quantum Langevin equation appears slightly different. It has been shown in \cite{bez,hanggi} that the occurrence of the initial slip term is actually due to the decoupled thermal initial state of the total system plus bath arrangement. Typically, the initial preparation of the total system-plus-bath model fixes the statistical properties of the bath operators as
well as the system degrees of freedom. Moreover, this initial preparation is the one which renders the fluctuating force ${\bf{F}}(t)$ a truly random one. In addition to this point, the force operator becomes stochastic only if the average of it is zero at all times (i.e., $\langle {\bf{F}}(t) \rangle = 0$). Also this quantum Brownian noise ${\bf{F}}(t)$ should constitute a stationary process with time-homogeneous correlations. But the average of ${\bf{F}}(t)$ with respect to Eq.(\ref{bath-density-matrix}) is a non zero quantity implying ${\bf{F}}(t)$ is non-Gaussian. But the force operator ${\bf{F}}(t)$ becomes a Gaussian random force once we take the average of it with respect to a bath density matrix which contains shifted oscillators. The initial preparation of the bath with shifted oscillators is given by the form
\begin{equation}
\hat{\rho}_B = \frac{1}{\mathcal{Z}}\exp\bigg\{\sum_{j=1} ^N
\bigg[\frac{{\bf{p}}_j ^2 (0)}{2m_j} + \frac{1}{2}m_j \omega_j ^2
  \bigg({\bf{x}}_j (0) - \frac{C_j {\bf{r}}(0)}{m_j \omega_j ^2}  \bigg)^2
  \bigg]  \bigg\} .
\end{equation}
Thus by absorbing the spurious slip term by a corresponding shift of the random force ${\bf{K}}(t)$, the force operator ${\bf{F}}(t)$ does reconcile with the usual Gaussian statistics as is required. 

The random force given by Eq.(\ref{friction}) satisfies the relations
\begin{subequations}
\begin{align}
\langle \{F_{\alpha} (t) , F_{\kappa} (t^{\prime}) \} \rangle &= \delta_{\alpha \kappa}\frac{2}{\pi}\int_0 ^\infty d\omega \Re[\tilde{\gamma}(\omega + i0^+)]\hbar\omega \coth\bigg(\frac{\beta\hbar\omega}{2}\bigg)\cos[\omega (t-t^{\prime})] ,\label{force-symmetric} \\
\langle [F_{\alpha} (t) , F_{\kappa} (t^{\prime}) ] \rangle &= \delta_{\alpha \kappa}\frac{2}{i \pi}\int_0 ^\infty d\omega \Re[\tilde{\gamma}(\omega + i0^+)]\hbar\omega \sin[\omega (t-t^{\prime})] ,\label{force-antisymmetric}
\end{align}
\end{subequations}
with $\alpha,\kappa = x,y,z$ and $\tilde{\gamma}(s) = \int_0 ^\infty dt \exp (ist)\gamma(t),~(\Im s >0)$. The angular brackets in the above equations show that we are thermal averaging over the heat bath. The fluctuating force ${\bf{F}}(t)$ lives in the Hilbert space of the heat bath. A heat bath is one which is capable of absorbing, through its many channels, energy infused by a system and this process is irreversible. In order that the Hamiltonian $\mathcal{H}_B$ (cf., Eq.(\ref{bath-ham})) to have the required properties of a heat bath, we must definitely go to a continuum by converting all the summations over $j$ to integrals over the frequency of the bath excitations by introducing a suitable spectral density of states $J(\omega)$ where
\begin{equation}
J(\omega) = \pi \sum_{j=1}^N \frac{C_j ^2}{2m_j \omega_j}\delta(\omega - \omega_j) ,
\label{spectral-density}
\end{equation}
so that
\begin{equation}
\gamma(t) = \sum_{j} \frac{C_j ^2}{m_j \omega_j ^2}\cos(\omega_j t) = 2 \int_0 ^\infty \frac{d\omega}{\pi}\frac{J(\omega)}{\omega}\cos(\omega t) .
\label{friction-density}
\end{equation}
In the framework of dissipative systems, the most employed spectral density is $J(\omega) =m\gamma \omega$, which is often referred as `Ohmic spectral density'. Note that this term is used sometimes to indicate a proportionality to frequency merely at low frequencies instead of the whole range of the frequency spectrum. In the ohmic case $\gamma (t-t^{\prime})$ is replaced by $m\gamma \delta (t-t^{\prime})$ so that $\Re [\tilde{\gamma}(\omega + i0^+)]$ reduces to a constant $m\gamma$. Also in the strict ohmic limit we recover the ordinary Langevin equation but surprisingly the underlying stochastic process described by the random force remains non-Markovian. In principle the spectral density cannot go to infinity for large values of $\omega$. So even if we introduce a cutoff to the frequency spectrum above which the spectral density vanishes, we can still use the term Ohmic for frequencies below the cutoff. But the spectral density $J(\omega)$ diverges in the strict ohmic limit for large frequencies and moreover the dispersion in the momentum shows an ultraviolet divergence while employing the ohmic spectrum. Therefore it is always advised to choose a cut off in the bath modes so that the divergences in the problem can be easily solved. Typically the \textit{Drude model} of spectral density is employed for the regularization purposes. When it comes to regularization of certain quantities it is assumed that $\omega_D ^{-1}$ is the shortest time scale available apart from the thermal time scale $\hbar \beta$, in the problem. For the Drude bath, the bath spectral density is assumed to have a smooth cutoff of the form
\begin{equation}
J(\omega) = m\gamma \omega \frac{\omega_D ^2}{\omega^2 + \omega_D ^2} ,
\end{equation}
which behaves like ohmic for small frequencies but will go to zero above the cut off $\omega_D$. We may observe from Eq.(\ref{friction-density}) that the damping kernel describes exponential memory on the timescale $\omega_D ^{-1}$. That means, for positive arguments $t>0$, the damping kernel takes a form
\begin{equation}
\gamma(t)=m\gamma\omega_D \exp(-\omega_D t) ,
\label{gamma-cutoff}
\end{equation}
with a damping strength $\gamma_0 = \int_0 ^\infty dt \gamma(t) = \gamma$, similar to ohmic one. If the timescale we are interested in is shorter than $\omega_D ^{-1}$ (usually this is the case when $\omega_D$ represents the largest frequency scale in the problem) the memory effects may often be neglected and the ohmic model may be employed instead. Fourier transform of Eq.(\ref{gamma-cutoff}) yields
\begin{equation}
\gamma(\omega) = \frac{m\gamma\omega_D}{\omega_D - i\omega},~~{\rm {and}}~\Re{[\frac{\gamma(\omega)}{m}]}=\frac{\gamma \omega_D ^2}{\omega_D ^2 + \omega^2}.
\end{equation}

With these remarks we organize our paper as follows: In sec.2 we present in detail the stochastic modeling of the system and in the subsequent subsections we derive the autocorrelation functions of the position and velocity. Discussions on the fluctuation-dissipation relation, classical diffusive behavior of the free charged particle in the limit of $\omega_0 \rightarrow 0$, and the relation between the dispersions and the reduced partition function, are included. Sec.3 is the conclusion and references follows thereafter.
\section{Stochastic modeling, position and momentum dispersions}
\label{sec:3}
We assume that the system and the bath are initially (at $t=0$) uncorrelated (uncoupled) to each other. This does mean that the initial coordinates of the system are not correlated with those of the heat bath, which immediately implies $\langle {\bf{r}}(0){\bf{F}}(t) \rangle = \langle \dot{{\bf{r}}}(0){\bf{F}}(t) \rangle = 0$. The system is said to be in an arbitrary non-equilibrium state initially and due to the interaction effects caused by the bath the system looses its energy eventually and it equilibrates with the bath. Obviously thermal equilibrium state is the one in which all the transients are completely died out and the equilibrium quantities are time-independent. Put it in another way, the system reaches its thermal equilibrium state once it completely ``explored" its phase-space. The heat bath is assumed to be in thermal equilibrium at a temperature $T$ so that the bath can be described by the usual Gibbsian canonical equilibrium density matrix $\rho_B = \exp (-\beta\mathcal{H}_B)/\mathcal{Z}_B$, where $\mathcal{H}_B$ is the bath Hamiltonian given by Eq.(\ref{bath-ham}), $\mathcal{Z}_B$ is the partition function of the bath and $\beta = 1/k_B T$ with $k_B$ being the Boltzmann constant. If we define the initial arbitrary density matrix of the system as $\rho_S (0)$, then we may write the total initial density matrix of the system-plus-bath as $\rho_T (t=0)=\rho_S (0)\times \rho_B (0)$. The interaction between the system and the bath is switched on at time $t=0^+$ and the system starts to evolve in the presence of the heat bath. The non-equilibrium quantum dynamical evolution of the system eventually reaches its steady state equilibrium (with the bath) asymptotically (as $t \rightarrow \infty$). In other words, the system reaches equilibrium with its heat bath over very large experimental or observation time scale. The system's characteristic decay time is much larger than the bath correlation time. This is a characteristic situation of quantum Brownian motion. This is in complete contrast to the case of a typical quantum optical situation where the systematic evolution of the reduced system is fast, which in turn means that coherent dynamics of the system goes through many cycles during a typical relaxation time.

By introducing the variables
\begin{equation}
z=x+iy, ~~F=F_x +iF_y,~~ {\rm{and}} ~~ \bar{\gamma}(t)=\frac{\gamma(t)}{m}+i\omega_c~,
\end{equation}
the Langevin equation given in Eq.(\ref{langevin1}) can be recast into a convenient form given by\cite{jishad2}
\begin{equation}
\label{langevin2}
 \ddot{z}+\int_0 ^t dt^{\prime}
\bar{\gamma}(t-t^{\prime})\dot{z}(t^{\prime})+\omega_0 ^2 z =\frac{F(t)}{m}~,
\end{equation}
where $\omega_c = e B/mc$ is the cyclotron frequency. We derived the quantum Langevin equation (cf. Eq.(\ref{langevin1})) (and thereby Eq.(\ref{langevin2})) from the Hamiltonian given in Eq.(\ref{hamiltonian}) by projecting out (or integrating out) the bath variables ${\bf{x}}_j$ and ${\bf{p}}_j$ from the full Hilbert space of the system-plus-bath. If at time $t=0$ the initial states of the system and the heat bath are uncorrelated and the bath is in thermal equilibrium at a temperature $T$ and described by a density matrix $\rho_B (0)$, then the reduced dynamics of the system for $t>0$ is completely determined by $\hat{\mathcal{H}}_S$. But the bare system Hamiltonian $\hat{\mathcal{H}}_S$ given in Eq.(\ref{sys-ham}) does not have any bath parameters or effects in it. Hence we need to modify the system Hamiltonian to incorporate the bath effects explicitly. Nevertheless, if we look at the quantum Langevin equation (Eqs.(\ref{langevin1}) and (\ref{langevin2})) we see that it is derived for the system's coordinates and the equation contains both the friction and random force. Which means the effects of the bath on the system, through the friction and the stochastic quantum Brownian noise, are included in the system's dynamical equation of motion. This clearly indicates that the system Hamiltonian (cf., Eq.(\ref{sys-ham})) is modified by the bath. We therefore rewrite the subsystem Hamiltonian as an effective stochastic Hamiltonian given by\cite{jishad2}
\begin{equation}
 \mathcal{H}_S ^{E}=\frac{1}{2}m\dot{z}\dot{z}^{\dagger}-\frac{1}{2}\hbar\omega_c
+\frac{1}{2}m\omega_0 ^2 zz^{\dagger}.
\label{ham-eff}
\end{equation}
The superscript ``$E$" stands for ``effective". The dependence of the Hamiltonian on the bath parameters is explicit from Eq.(\ref{langevin2}). From Eq.(\ref{ham-eff}) the mean squared average values of ${\bf{r}}$ and $({\bf{p}}-e{\bf{A}}/c)$ can be determined from the two equal time correlation functions 
\begin{equation}
 \mathcal{C}(t)=\langle z(t)z^{\dagger}(t) \rangle ,~~{\rm and}~~\mathcal{D} (t)=\langle \dot{z}(t)\dot{z}^{\dagger}(t) \rangle.
\label{equal1}
\end{equation}
We start with the unequal time correlation functions 
\begin{equation}
\mathcal{C}(t,t^{\prime})=\langle z(t)z^{\dagger}(t^{\prime}) \rangle ,~~~ \mathcal{D} (t,t^{\prime})=\langle \dot{z}(t)\dot{z}^{\dagger}(t^{\prime}) \rangle,
\end{equation}
and then do an analytical continuation $t^{\prime}=t$ to obtain Eq.(\ref{equal1}). In order to find the solution of the Langevin equation in Eq.(\ref{langevin2}), we here introduce the fundamental solutions $\mathcal{G}_+ (t)$ and $\mathcal{G}_- (t)$ of the homogeneous part of the Langevin equation, which is basically obtained by setting the right hand side equal to zero. The fundamental solutions are usually defined through the initial conditions given by
\begin{eqnarray}
\mathcal{G}_+ (0) &=& 1,~~~\dot{\mathcal{G}}_+ (0)=0 , \\
\mathcal{G}_- (0) &=& 0,~~~\dot{\mathcal{G}}_- (0)=1 .
\end{eqnarray}
Making use of the Laplace transform
\begin{equation}
\hat{f}(s)=\int_0 ^\infty dt e^{-st} f(t) ,
\end{equation}
we can now write the Laplace transform of the fundamental solutions as follows
\begin{eqnarray}
\hat{\mathcal{G}}_+ (s) &=& \frac{s}{s^2 + \omega_0 ^2 + s\hat{\bar{\gamma}}(s)} , \\
\hat{\mathcal{G}}_- (s) &=& \frac{1}{s^2 + \omega_0 ^2 + s\hat{\bar{\gamma}}(s)}.\label{second-green}
\end{eqnarray}
Here $\hat{\bar{\gamma}}(s)$ is the Laplace transform of the damping kernel. Now with the help of the fundamental solutions, we can easily write the general solution of the quantum Langevin equation which is given by
\begin{equation}
z(t)=\mathcal{G}_+ (t) z(0) + \mathcal{G}_- (t) \dot{z}(0) +\frac{1}{m} \int_0 ^t d\tau \mathcal{G}_- (t-\tau)F(\tau) .
\label{langevin-green}
\end{equation}
It is easy to find out $\mathcal{G}_- (t)$. An inverse Laplace transform, with Eq.(\ref{second-green}), gives
\begin{equation}
\mathcal{G}_- (t) = \frac{1}{2\pi i}\int_{-i\infty} ^{+i \infty} \frac{e^{st}}{s^2 + \omega_0 ^2 + s\hat{\bar{\gamma}}(s)}ds = \frac{1}{2\pi i}\int_{-i\infty} ^{+i \infty} \frac{e^{st}}{(s+\mu_1)(s+\mu_2)}ds .
\label{green2}
\end{equation}
Taking a derivative of $\mathcal{G}_- (t)$ with respect to $t$ in the above equation, we find
\begin{equation}
\dot{\mathcal{G}}_- (t) =  \mathcal{G}_+ (t) ,
\end{equation}
so that the general solution given in Eq.(\ref{langevin-green}) can be written in terms of $\mathcal{G}_- (t)$ only and is given by
\begin{equation}
z(t)=\dot{\mathcal{G}}_- (t) z(0) + \mathcal{G}_- (t) \dot{z}(0) +\frac{1}{m} \int_0 ^t ds \mathcal{G}_- (t-\tau)F(\tau) .
\label{langevin-solution-green}
\end{equation}
From the formal solution given in terms of $\mathcal{G}_- (t)$, one must notice that we have used the homogeneous equation
\begin{equation}
 \ddot{\mathcal{G}}_- (t) +\int_0 ^t dt^{\prime}
\bar{\gamma}(t-t^{\prime})\dot{\mathcal{G}}_- (t^{\prime})+\omega_0 ^2 \mathcal{G}_- (t) = 0.
\end{equation}
After evaluating the integral by contour integration method, from Eq.(\ref{green2}), we obtain
\begin{equation}
\mathcal{G}_- (t) = -\frac{1}{(\mu_1 - \mu_2)}[e^{-\mu_1 t}-e^{-\mu_2 t}] ,
\label{green-second}
\end{equation}
with $\mu_{1,2} = -\frac{\hat{\bar{\gamma}}(s)}{2} \pm \sqrt{\frac{\hat{\bar{\gamma}} ^2 (s)}{4}-\omega_0 ^2}$. Therefore the final solution to the quantum Langevin equation can be obtained by substituting the value of $\mathcal{G}_- (t)$ in the solution given in Eq.(\ref{langevin-solution-green}). We point out that the Green's function $\mathcal{G}_- (t)$ introduced in this section is nothing but the response function $\chi(t)$ of the system (cf.Eq.(\ref{response-stationary})) to an external force. Therefore in our following discussions we tacitly write the response function $\chi(t)$ instead of $\mathcal{G}_- (t)$. This argument can be made more clear by changing (analytic continuation) the Laplace transform
\begin{equation}
\mathcal{G}_- (t-\tau) = \frac{1}{2\pi i}\int_{-i\infty} ^{+i \infty} \hat{\mathcal{G}}_- (s) e^{s(t-\tau)}ds ,
\end{equation}
to the Fourier (frequency) domain by the substitution $s=-i\omega$ and using the property of the memory friction function $\hat{\gamma}(s)=\tilde{\gamma}(\omega=is)$ and $\tilde{\gamma}(\omega) = \lim_{\epsilon \rightarrow 0^+}\hat{\gamma}(s=-i\omega+\epsilon)$. We write
\begin{equation}
\mathcal{G}_- (t-\tau) = \frac{1}{2\pi}\int_{-\infty} ^{+ \infty} d\omega \mathcal{G}_- (-i\omega) e^{-i\omega(t-\tau)} ,
\end{equation}
where (cf., Eq.(\ref{second-green}))
\begin{equation}
\mathcal{G}_- (-i\omega) = \frac{1}{(-\omega^2 -i\omega \bar{\gamma}(\omega)+\omega_0 ^2)}~~;~~ \bar{\gamma}(\omega)=i\omega_c +\frac{\tilde{\gamma}(\omega)}{m} .
\end{equation}
Note that the right hand side of the above equation is nothing but the dynamical susceptibility which is easily obtained by taking a Fourier transform of the Langevin equation given in Eq.(\ref{langevin2}). The relation connecting the response function and the dynamical susceptibility can be expressed as
\begin{equation}
\chi(t-\tau)=\frac{1}{2\pi}\int_{-\infty}^ {+\infty} \chi(\omega)e^{-i\omega(t-\tau)} d\omega = \frac{1}{2\pi}\int_{-\infty}^ {+\infty} \frac{1}{(-\omega^2 -i\omega \bar{\gamma}(\omega)+\omega_0 ^2)} e^{-i\omega(t-\tau)} d\omega ,
\label{response-reference-function}
\end{equation}
which clearly shows us that the Green's function $\mathcal{G}_- (t)$ is indeed the response function $\chi(t)$.

The causal response function $\chi(t)$ is called the generalized susceptibility. The force $F(t)$ on the right hand side of the Langevin equation in Eq.(\ref{langevin2}) satisfies the relations Eq.(\ref{force-symmetric}) and Eq.(\ref{force-antisymmetric}). Typically the memory friction $\gamma(t)$ falls to zero in the bath relaxation time which is very short compared to the system's time scale. For times greater than the bath relaxation time, the initial slippage term in the Langevin equation vanishes. That means the spurious initial term vanishes for very long times which is compared to the system's characteristic decay time. Therefore for very long times compared to the system's time scale, the Langevin equation becomes a stationary one with the lower limit of the integration being $-\infty$. i.e.,
\begin{equation}
\label{langevin3}
 \ddot{z}+\int_{-\infty} ^t dt^{\prime}
\bar{\gamma}(t-t^{\prime})\dot{z}(t^{\prime})+\omega_0 ^2 z =\frac{F(t)}{m} ,
\end{equation}
for which we usually write the solution
\begin{equation}
 z_S (t)=\int_{-\infty} ^t d\tau \chi(t-\tau)\frac{F(\tau)}{m} ,
\label{response-stationary}
\end{equation}
where
\begin{equation}
\chi(t-\tau)=\frac{1}{2\pi}\int_{-\infty}^ {+\infty} \chi(\omega)e^{-i\omega(t-\tau)} d\omega ,
\label{green function}
\end{equation}
where $\chi(\omega)$ which is called the dynamical susceptibility can be deduced from Eq.(\ref{langevin2}) or Eq.(\ref{langevin3}), and is given by 
\begin{equation}
 \chi (\omega)=\frac{1}{(-\omega^2 -i\omega \bar{\gamma}(\omega)+\omega_0
^2)} ,
\label{laplace}
\end{equation}
where
\begin{equation}
\bar{\gamma}(\omega)=i\omega_c +\frac{\tilde{\gamma}(\omega)}{m}.
\label{add-up}
\end{equation}
The expression for the dynamical susceptibility in Eq.(\ref{laplace}) is identical to the corresponding one for a classical damped charged oscillator in a magnetic field with frequency dependent damping. The absence of any quantum correction to the dynamical susceptibility is actually a consequence of the Ehrenfest's theorem for strictly linear systems like what we consider here. The force $F(t)$ in Eq.(\ref{langevin3}) satisfies the same relations given in Eqs.(\ref{force-symmetric}) and (\ref{force-antisymmetric}). The subscript $S$ indicates that $z_S (t)$ is a stationary process, in the sense that all the correlations and probability distributions for this dynamical variable $z_S (t)$ are time translational invariant ($t\rightarrow t+t_0$). Moreover the symmetrized part of the position autocorrelation is a function of the difference of two different times only. From Eq.(\ref{green function}) it is very much clear that so long as the frequency $\omega$ remains non zero, the response function $\chi(t)$ will vanish exponentially for much longer times. This is in accordance with the Tauberian theorem which says that the asymptotic behavior of a function depends upon the low frequency behavior of the Fourier transform of the function. These principles imply that, for long times, the dependence of $z(t)$ on the initial coordinates in Eq.(\ref{langevin-solution-green}) disappears completely (it is very much visible from the solution in Eq.(\ref{langevin-solution-green}) with Eq.(\ref{green-second}) that for long time $t\rightarrow \infty$ the terms containing the initial values $z(0)$ and $\dot{z}(0)$ vanish) and we write 
\begin{equation}
 z(t)=\int_0 ^t d\tau \chi(t-\tau)\frac{F(\tau)}{m} .
\label{response}
\end{equation}
Comparing Eq.(\ref{response}) with Eq.(\ref{response-stationary}) for $z_S (t)$ implies that, $z(t)$ in Eq.(\ref{response}) becomes the solution of the stationary Langevin equation given in Eq.(\ref{langevin3}). Fourier transform of $z(t)$ is given by $\tilde{z}(\omega)=\tilde{\chi}(\omega)\tilde{F}(\omega)/m$, where $\tilde{F}(\omega)$ is the Fourier transform of the force $F(\tau)$. Taking a Laplace transform, rearranging terms and then an inverse Laplace transform leads Eq.(\ref{response}) to have the explicit form
\begin{equation}
z(t) = \frac{1}{m}\int_0 ^t e^{-\frac{\bar{\gamma}}{2}(t-\tau)}\frac{\sinh\bigg[\sqrt{\omega_0 ^2 - \frac{\bar{\gamma}^2}{4}} (t-\tau)\bigg]}{\sqrt{\omega_0 ^2 - \frac{\bar{\gamma}^2}{4}}}F(\tau)d\tau .
\end{equation}

For the Drude bath the dynamical susceptibility $\chi(\omega)$ can be written as 
\begin{equation}
\chi (\omega)=\frac{1}{(-\omega^2 -i\omega (i\omega_c + \frac{\gamma\omega_D}{\omega_D -i\omega})+\omega_0^2)} .
\label{chiomega}
\end{equation}
Alternatively,
\begin{equation}
\label{chiomega-roots}
 \chi(\omega)=-\frac{(\omega+i\omega_D)}{
(\omega+i\lambda_1)(\omega+i\lambda_2)(\omega+i\lambda_3)} ,
\end{equation}
where $\lambda_j s$ are the roots of the cubic equation in the denominator of Eq.(\ref{chiomega}), and the roots satisfy the vieta equations\cite{jishad1} 
\begin{equation}
\begin{split}
\lambda_1 + \lambda_2 + \lambda_3 &= \omega_D + i\omega_c ,\\
\lambda_1 \lambda_2 + \lambda_2 \lambda_3 + \lambda_3 \lambda_1 &= \omega_0 ^2 + \gamma\omega_D + i\omega_c \omega_D ,\\
\lambda_1 \lambda_2 \lambda_3 &= \omega_0 ^2 \omega_D .
\end{split}
\label{vieta}
\end{equation}
Similarly we can write $\chi^* (\omega)$ as the complex conjugate of Eq.(\ref{chiomega-roots}) and is given by
\begin{equation}
\label{chiomega-roots1}
 \chi^* (\omega)=-\frac{(\omega-i\omega_D)}{
(\omega-i\lambda_1 ^{\prime})(\omega-i\lambda_2 ^{\prime})(\omega-i\lambda_3 ^{\prime})} ,
\end{equation}
where the $\lambda^{\prime}_j s$ are the complex conjugates of $\lambda$'s and satisfy the vieta equations
\begin{equation}
\begin{split}
\lambda_1 ^{\prime} + \lambda_2 ^{\prime} + \lambda_3 ^{\prime} &= \omega_D - i\omega_c ,\\
\lambda_1 ^{\prime} \lambda_2 ^{\prime} + \lambda_2 ^{\prime} \lambda_3 ^{\prime} + \lambda_3 ^{\prime} \lambda_1 ^{\prime} &= \omega_0 ^2 + \gamma\omega_D - i\omega_c \omega_D ,\\
\lambda_1 ^{\prime} \lambda_2 ^{\prime} \lambda_3 ^{\prime} &= \omega_0 ^2 \omega_D .
\end{split}
\label{vieta1}
\end{equation}
In the absence of the magnetic field ($\omega_c = 0$), the vieta equations are exactly matching with the corresponding equations for a damped quantum harmonic oscillator. Moreover, in the absence of the magnetic field, the vieta equations for $\lambda_j$'s and $\lambda^{\prime} _j$'s are equivalent. 
 \subsection{Position autocorrelation function}
\label{sec:4}
In this section we thoroughly study the autocorrelation function of the position of a quantum charged oscillator in the presence of a magnetic field and a heat bath. The correlation function $\mathcal{C} (t,t^{\prime}) = \langle z(t)z^{\dagger} (t^{\prime}) \rangle$ is evaluated here. Using $z=x+iy$, the real and the imaginary parts of $\mathcal{C} (t,t^{\prime}) = \langle z(t)z^{\dagger} (t^{\prime}) \rangle$ can be expressed as
\begin{equation}
z(t)z^{\dagger} (t^{\prime}) = [x(t)x(t^{\prime}) + y(t)y(t^{\prime})] + i[y(t)x(t^{\prime})-x(t)y(t^{\prime})] .
\label{real and imaginary}
\end{equation}
Therefore in the limit $t^{\prime} = t$ the real part of the correlation contributes to $ \mathcal{C} (t,t^{\prime})$, which in turn gives the mean-squared displacement $\langle z(t)z^{\dagger} (t^{\prime}) \rangle = \langle x^2 (t) + y^2 (t) \rangle$ of the particle in the $xy$-plane, and the imaginary part vanishes. We can also write the correlation function $\mathcal{C}(t,t^{\prime}) = \langle z(t)z^{\dagger} (t^{\prime}) \rangle$ in terms of the symmetric combination and the commutator of $z(t)$ and $z^{\dagger} (t^{\prime})$ in the following way
\begin{equation}
\mathcal{C} (t,t^{\prime}) = \langle z(t)z^{\dagger} (t^{\prime}) \rangle = \mathcal{S}(t,t^{\prime})+i\mathcal{A}(t,t^{\prime}),
\end{equation}
where
\begin{equation}
\mathcal{S}(t,t^{\prime}) = \frac{1}{2}\langle \{z(t), z^{\dagger} (t^{\prime}) \} \rangle,~~{\rm and}~~\mathcal{A}(t,t^{\prime}) = \frac{1}{2i} \langle [z(t), z^{\dagger} (t^{\prime}) ] \rangle .
\end{equation}
Again at equal times ($t^{\prime}=t$) the symmetrized part does contribute to give the mean squared displacement $\langle x^2 (t) + y^2 (t) \rangle$ and the commutator structure vanishes. Using Eq.(\ref{response}), the symmetrized part of the position correlation can be easily written as
\begin{equation}
\mathcal{S}(t,t^{\prime}) = \frac{1}{2} \langle \{z(t), z^{\dagger} (t^{\prime}) \} \rangle = \frac{1}{2 m^2} \int_0 ^t d\tau \int_0 ^{t^{\prime}} d\tau^{\prime}
\chi(t-\tau)\chi^* (t^{\prime} -\tau^{\prime}) \langle \{ F(\tau), F^{\dagger}(\tau^{\prime}) \} \rangle ,
\label{pos-corr-symmetric}
\end{equation}
where, using (\ref{force-symmetric}), we can write
\begin{eqnarray}
\langle \{ F(\tau), F^{\dagger}(\tau^{\prime}) \} \rangle &=& \langle \{F_x (\tau), F_x (\tau^{\prime})  \}\rangle + \langle \{F_y (\tau), F_y (\tau^{\prime})  \}\rangle , \nonumber\\
&=& \frac{4}{\pi}\int_0 ^\infty d\tilde{\omega} \Re[\tilde{\gamma}(\tilde{\omega} + i0^+)]\hbar\tilde{\omega} \coth\bigg(\frac{\beta\hbar\tilde{\omega}}{2}\bigg)\cos[\tilde{\omega} (\tau-\tau^{\prime})] .
\label{force-symmetric1}
\end{eqnarray}
Now using Eq.(\ref{green function}) in Eq.(\ref{pos-corr-symmetric}) and with the cognizance of the evenness of the integrand in the Eq.(\ref{force-symmetric1}) we can now write Eq.(\ref{pos-corr-symmetric}), after evaluating the integrals over $\tau$ and $\tau^\prime$, as
\begin{eqnarray}
\label{zeta-modified-1}
\mathcal{S}(t,t^{\prime}) &=& \frac{1}{4\pi^3 m^2} \int_{-\infty}^{+\infty}d\tilde{\omega} \frac{m\gamma \omega_D ^2}{(\omega_D ^2 +\tilde{\omega}^2)} \hbar\tilde{\omega} \coth \bigg(\frac{\beta\hbar\tilde{\omega}}{2} \bigg)     \int_{-\infty}^{+\infty}d\omega
\chi(\omega)\frac{(e^{-i\tilde{\omega}t}-e^{-i\omega
t})}{i(\omega-\tilde{\omega})}\nonumber\\
&\times& \int_{-\infty}^{+\infty}d\omega^{\prime}\chi^*
(\omega^{\prime})\frac{(e^{i\tilde{\omega}t^{\prime}}-e^{i\omega^{\prime}
t^{\prime}})}{-i(\omega^{\prime}-\tilde{\omega})} .
\end{eqnarray}
After evaluating the last two integrals, we obtain
\begin{equation}
\label{zeta-trignometric}
\mathcal{S}(t,t^{\prime}) = \frac{1}{4\pi^3 m} \int_{-\infty}^{+\infty} d\tilde{\omega} \frac{\gamma \omega_D ^2}{(\omega_D ^2 +\tilde{\omega}^2)} \hbar\tilde{\omega} \coth \bigg(\frac{\beta\hbar\tilde{\omega}}{2} \bigg)\mathcal{K}_1 (\tilde{\omega},\omega_D,\lambda_j , t)\mathcal{K}_2 (\tilde{\omega},\omega_D,\lambda^{\prime} _j , t^{\prime}) ,
\end{equation}
where
\begin{equation}
\begin{split}
\mathcal{K}_1 (\tilde{\omega},\omega_D,\lambda_j , t) &= \frac{2\pi}{i \mathcal{M}} \bigg\{\frac{(\lambda_1 -\omega_D)(\lambda_2
-\lambda_3)(e^{-i\tilde{\omega}t}-e^{-\lambda_1
t})}{(\tilde{\omega}+i\lambda_1)} + \frac{(\lambda_2 -\omega_D)(\lambda_3
-\lambda_1)(e^{-i\tilde{\omega}t}-e^{-\lambda_2
t})}{(\tilde{\omega}+i\lambda_2)} \\
&+ \frac{(\lambda_3 -\omega_D)(\lambda_1
-\lambda_2)(e^{-i\tilde{\omega}t}-e^{-\lambda_3
t})}{(\tilde{\omega}+i\lambda_3)}\bigg\},
\label{integrals-modified}
\end{split}
\end{equation}
and
\begin{equation}
\begin{split}
\mathcal{K}_2 (\tilde{\omega},\omega_D,\lambda^{\prime} _j , t^{\prime}) &= -\frac{2\pi}{i \mathcal{M}^{\prime}} \bigg\{ \frac{(\lambda^{\prime} _{1}
-\omega_D)(\lambda^{\prime}_2
-\lambda^{\prime}_3)(e^{i\tilde{\omega}t^{\prime}}-e^{-\lambda^{\prime}_1
t^{\prime}})}{(\tilde{\omega}-i\lambda^{\prime}_1)} 
+ \frac{(\lambda^{\prime}_2 -\omega_D)(\lambda^{\prime}_3
-\lambda^{\prime}_1)(e^{i\tilde{\omega}t^{\prime}}-e^{-\lambda^{\prime}_2
t^{\prime}})}{(\tilde{\omega}-i\lambda^{\prime}_2)} \\
&+ \frac{(\lambda^{\prime}_3 -\omega_D)(\lambda^{\prime}_1
-\lambda^{\prime}_2)(e^{i\tilde{\omega}t^{\prime}}-e^{-\lambda^{\prime}_3
t^{\prime}})}{(\tilde{\omega}-i\lambda^{\prime}_3)}\bigg\} .
\label{integrals-modified-1}
\end{split}
\end{equation}
Here
\begin{eqnarray}
\label{integral-constants}
\mathcal{M}&=&(\lambda_1 -\lambda_2)(\lambda_1 -\lambda_3)(\lambda_2 -\lambda_3) ,\\
\mathcal{M}^{\prime}&=&(\lambda^{\prime}_1 -\lambda^{\prime} _2)(\lambda^{\prime}_1
-\lambda^{\prime}_3)(\lambda^{\prime}_2 -\lambda^{\prime}_3) .
\end{eqnarray}

The integral in Eq.(\ref{zeta-trignometric}) contains transient terms which depend on times $t$ and $t^{\prime}$, but not their difference explicitly. If we consider the times $t$ and $t^{\prime}$ to be very large ($t,t^{\prime} \rightarrow \infty$), but not their difference $t-t^{\prime}$, then all the transients vanish. In this case we write
\begin{eqnarray}
\label{zeta-steady-state}
\mathcal{S}(t,t^{\prime}) &=& \frac{\hbar}{4 \pi^3 m}\int_{-\infty}^{+\infty}
d\tilde{\omega}\frac{\gamma \omega_D ^2}{(\omega_D ^2
+\tilde{\omega}^2)}\tilde{\omega} \coth \bigg(\frac{\beta\hbar\tilde{\omega}}{2} \bigg) e^{-i\tilde{\omega} (t-t^{\prime})}\nonumber\\
&\times& \bigg(\frac{2\pi}{i} \frac{(\omega_D
-i\tilde{\omega})}{(\tilde{\omega}+i\lambda_1)(\tilde{
\omega}+i\lambda_2)(\tilde{\omega}+i\lambda_3)}\bigg) \nonumber\\
&\times& \bigg(-\frac{2\pi}{i} \frac{(\omega_D
+i\tilde{\omega})}{(\tilde{\omega}-i\lambda^{\prime}
_1)(\tilde{\omega}-i\lambda^{\prime}_2)(\tilde{\omega}-i\lambda^{\prime}_3)} \bigg) ,
\end{eqnarray}
which, by using the definitions of the susceptibilities and the fact that $\Re[{\tilde{\gamma}(\tilde{\omega})}/m]=\gamma\omega_D ^2 / (\omega_D ^2 + \tilde{\omega}^2)$, becomes
\begin{equation}
\mathcal{S}(t,t^{\prime}) = \frac{\hbar}{m \pi}\int_{-\infty}^{+\infty}
d\tilde{\omega}~\tilde{\omega}\Re[\frac{\tilde{\gamma}(\tilde{\omega})}{m}]\chi(\tilde{\omega})\chi^* (\tilde{\omega}) \coth \bigg(\frac{\beta\hbar\tilde{\omega}}{2} \bigg) e^{-i\tilde{\omega} (t-t^{\prime})} .
\label{symmetric-fdt}
\end{equation}
From the definitions of $\chi(\omega)$ and $\chi^* (\omega)$, we can easily show the following relation
\begin{equation}
\tilde{\omega}\Re[\frac{\tilde{\gamma}(\tilde{\omega})}{m}]\chi(\tilde{\omega})\chi^* (\tilde{\omega}) = \frac{1}{2i}[\chi(\tilde{\omega}) - \chi^* (\tilde{\omega})],
\label{separation}
\end{equation}
using which evaluation of the integral in Eq.(\ref{symmetric-fdt}) is trivial. The symmetric part $\mathcal{S}(t,t^{\prime})$ of the autocorrelation function $\mathcal{C}(t,t^{\prime})$ shows more interesting features. Clearly one can see the emergence of two types of times scales which are determined by the poles of the integrand in Eq.(\ref{zeta-steady-state}). One set of poles are from the $\lambda_j$'s and the other set of poles are due to the Matsubara frequencies $\nu_n$ which originates from the infinite sequence of simple poles of the $\coth(\beta\hbar\tilde{\omega} /2)$ located at $\tilde{\omega} = \pm i\nu_n$, where $n=1,2,3,....~$. In view of the two contributions, we express the symmetrized part of the autocorrelation function in terms of two components, ie.,
\begin{equation}
\mathcal{S}(t)=S_1 (t)+S_2 (t) ,
\end{equation}
where $S_1 (t)$ is due to the poles determined by the roots $\lambda_j$s and $S_2 (t)$ is due to the poles at Matsubara frequencies. Using Eq.(\ref{separation}) in Eq.(\ref{symmetric-fdt}) and after a contour integration, we obtain
\begin{eqnarray}
S_1 (t) &=& \frac{i\hbar}{m}\bigg\{\coth\bigg(\frac{i\beta\hbar\lambda_1}{2} \bigg)\frac{(\omega_D -\lambda_1)}{(\lambda_1 -\lambda_2)(\lambda_1 -\lambda_3)}e^{-\lambda_1 t}- \coth\bigg(\frac{i\beta\hbar\lambda_2}{2} \bigg)\frac{(\omega_D -\lambda_2)}{(\lambda_1 -\lambda_2)(\lambda_2 -\lambda_3)}e^{-\lambda_2 t} \nonumber\\
&+& \coth\bigg(\frac{i\beta\hbar\lambda_3}{2} \bigg)\frac{(\omega_D -\lambda_3)}{(\lambda_1 -\lambda_3)(\lambda_2 -\lambda_3)}e^{-\lambda_3 t} \bigg\} ,
\end{eqnarray}
which, in the strict ohmic limit ($\omega_D \rightarrow \infty$), becomes
\begin{equation}
S_1 (t) = \frac{\hbar}{im(\lambda_2 -\lambda_1)}\bigg\{\coth\bigg(\frac{i\beta\hbar\lambda_2}{2}\bigg)e^{-\lambda_2 t} - \coth\bigg(\frac{i\beta\hbar\lambda_1}{2}\bigg)e^{-\lambda_1 t}  \bigg\} ,
\end{equation}
which matches with the corresponding result for a damped quantum oscillator in two dimensions\cite{weissbook}. We have assumed here that in the limit of $\omega_D \rightarrow \infty$, one of the roots, let us say, $\lambda_3 \approx \omega_D$ and $\lambda_{1,2} = \bar{\gamma}/2 \pm 1/2 \sqrt{\bar{\gamma}^2 -4\omega_0 ^2}$. Also in the limit of $\bar{\gamma} \ll 2\omega_0$, we write $\lambda_{1,2} = \bar{\gamma}/2 \pm i \sqrt{\omega_0 ^2 - \frac{\bar{\gamma}^2}{4}}$, and that yields
\begin{equation}
S_1 (t) = \frac{\hbar}{m\Omega}e^{-\frac{\bar{\gamma}}{2}t}\bigg\{\frac{\sinh(\beta\hbar\Omega)\cos(\Omega t) + \sinh(\beta\hbar\bar{\gamma} / 2)\sin(\Omega t)}{\cosh(\beta\hbar\Omega)-\cos(\beta\hbar\bar{\gamma} / 2)}  \bigg\} ,
\end{equation}
where $\Omega = \sqrt{\omega_0 ^2 - \bar{\gamma}^2 /4}$. For $\bar{\gamma} \gg 2\omega_0$, $S_1 (t) \sim -\hbar/im\bar{\gamma}\coth(i\beta\hbar\omega_0 ^2 /2\bar{\gamma})e^{-\omega_0 ^2 t /\bar{\gamma}}$, and for the high temperatures $k_B T \gg \hbar\omega_0 ^2 /\bar{\gamma}$, it becomes $(2/m\beta\omega_0 ^2) e^{-\omega_0 ^2 t /\bar{\gamma}}$. The other contribution $S_2 (t)$, emanates from the infinite sequence of simple poles of $\coth(\beta\hbar\tilde{\omega} /2)$ located at $\tilde{\omega}  = \pm i\nu_n ~ (n=1,2,3...)$, is given by
\begin{equation}
S_2 (t) = \frac{2}{m\beta}\sum_{n=1}^\infty \bigg\{\frac{(\omega_D - \nu_n)}{(\nu_n - \lambda_1)(\nu_n - \lambda_2)(\nu_n - \lambda_3)}e^{-\nu_n t}+ \frac{(\omega_D + \nu_n)}{(\nu_n + \lambda_1 ^{\prime})(\nu_n + \lambda_2 ^{\prime})(\nu_n + \lambda_3 ^{\prime})}e^{-\nu_n t}  \bigg\} .
\label{second-correlation}
\end{equation}
In the strict ohmic case, $S_2 (t)$ can be written as
\begin{equation}
S_2 (t) = -\frac{4\gamma}{m\beta}\sum_{n=1}^\infty \frac{\nu_n e^{-\nu_n t}}{(\nu_n ^2 + \omega_0 ^2)^2 -2i\omega_c \nu_n (\nu_n ^2 + \omega_0 ^2)-\nu_n ^2 (\gamma^2 + \omega_c ^2)} .
\label{second-correlation-nocutoff}
\end{equation}
In the absence of the magnetic field ($\omega_c  =0$), from Eq.(\ref{second-correlation-nocutoff}), we get
\begin{equation}
S_2 (t) = -\frac{4\gamma}{m\beta}\sum_{n=1}^\infty \frac{\nu_n e^{-\nu_n t}}{(\nu_n ^2 + \omega_0 ^2)^2 -\nu_n ^2 \gamma^2} ,
\end{equation}
which is matching with the corresponding result of a quantum damped oscillator in two dimensions. 

At high temperatures (or for $\hbar \rightarrow 0$), the contribution from $S_2 (t)$ is very small and its drops to zero much faster than $S_1 (t)$. Therefore, for the symmetric part of the autocorrelation function of the position, we have
\begin{equation}
\mathcal{S}(t) = \frac{2}{m\beta \omega_0 ^2} \{\cos(\Omega t)+\frac{\bar{\gamma}}{2\Omega}\sin(\Omega t) \}e^{-\bar{\gamma}|t|/2} ,
\label{classical-correlation}
\end{equation}
which for $t=0$ results in
\begin{equation}
\mathcal{S}(0) = \frac{2}{m\beta \omega_0 ^2} ,
\end{equation}
the expected classical result. Thus we have clearly elucidated the classical behavior of the autocorrelation function (cf., Eq.(\ref{classical-correlation})) in the appropriate classical limit. However, $S_2 (t)$ becomes seemingly important at very low-temperatures where the quantum effects are predominant. To discuss the quantum effects we focus our attention on the zero temperature case where we see the effect of the zero point oscillations. In this case the Matsubara frequencies $\nu_n$ become dense and get closer to each other and at $T=0$ all of them contributes. In this situation we replace the summation by an integral such that
\begin{equation}
\mathcal{S}(t) = -\frac{2\hbar\gamma}{m\pi}\int_0 ^\infty dx \frac{x e^{-xt}}{(x^2 +\omega_0^2 -\bar{\gamma}x)(x^2 +\omega_0^2 +\bar{\gamma}^* x)}.
\end{equation}
This integral may be expressed in terms of linear combination of exponential integral functions. The long time behavior, i.e., $t > 1/\omega_0$, $\bar{\gamma} / \omega_0 ^2$, $\bar{\gamma}^* / \omega_0 ^2$, may be easily obtained asymptotically by replacing the integrand by its behavior for small values of the integration variable. This results in the algebraic decay in time with the leading order term,
\begin{equation}
\mathcal{S}(t) \sim -\frac{2\hbar\gamma}{m\pi}\frac{1}{\omega_0 ^4} \int_0 ^\infty dx~ x~ e^{-xt} = -\frac{2\hbar\gamma}{m\pi}\frac{1}{\omega_0 ^4} \frac{1}{t^2} .
\label{algebraic}
\end{equation}
It is to be noted here that $\mathcal{S}(t)$ can be evaluated, at $T=0$, from Eq.(\ref{second-correlation}) as well in the presence of the Drude cutoff. However the algebraic decay remains unaffected by the presence of the Drude cutoff. This in turn means the Drude cutoff does not change the low frequency behavior of the spectral density of the bath oscillators. The algebraic $1/t^2$ power law decay at long times for the correlation function is a characteristic feature of the ohmic damping. Interestingly the magnetic field has no effect on the long time behavior of the symmetrized correlation function at zero temperature. 

However, at finite temperatures the long time power law decay in time is replaced by an exponential decay. For $\nu t \gg 1$, one can replace the sum in Eq.(\ref{second-correlation-nocutoff}) by its first term ($n=1$). This essentially gives, for the position autocorrelation at long times,
\begin{equation}
\mathcal{S}(t) = -\frac{4\gamma}{m\beta} \frac{\nu e^{-\nu t}}{(\nu^2 + \omega_0 ^2)^2 -2i\omega_c \nu(\nu^2 + \omega_0 ^2)-\nu^2 (\gamma^2 + \omega_c ^2)} .
\label{long-time-finite-temperature}
\end{equation}
Here $\nu = 2\pi / \hbar\beta$ is the first bosonic Matsubara frequency. In a sense, $\mathcal{S}(t) \propto e^{-\nu t}$. The qualitative change we observe in the long time behavior of the position autocorrelation, from a power-law decay at zero temperature ($T=0$) to an exponential decay at any finite temperature ($T > 0$), is a typical quantum mechanical phenomenon. The antisymmetric part of the position autocorrelation function never appears in these discussions since that part never contributes anything to these long time effects at zero and finite temperatures as it does not contain any temperature dependent factor in it. But we discuss it here in the next paragraph for the completeness.

The antisymmetric part $\mathcal{A}(t,t^\prime)$ of the position autocorrelation function involving the commutator structure is defined earlier as $\mathcal{A}(t,t^\prime) = \frac{1}{2i}\langle [z(t),z^\dagger (t^{\prime})] \rangle$.  $\mathcal{A}(t,t^\prime)$ can be calculated using the commutator structure of the force-force correlation in Eq.(\ref{force-antisymmetric}) as
\begin{eqnarray}
\mathcal{A}(t,t^\prime) = \frac{1}{2i}\langle [z(t),z^\dagger (t^{\prime})] \rangle &=& \frac{1}{2i m^2} \int_0 ^t d\tau \int_0 ^{t^{\prime}} d\tau^{\prime} \chi(t-\tau)\chi^* (t^{\prime} -\tau^{\prime}) \langle [ F(\tau), F^{\dagger}(\tau^{\prime}) ] \rangle \nonumber\\
&=& \frac{1}{2 i m^2} \int_0 ^t d\tau \int_0 ^{t^{\prime}} d\tau^{\prime} \chi(t-\tau)\chi^* (t^{\prime} -\tau^{\prime}) \nonumber\\
&\times& \frac{4}{i \pi}\int_0 ^\infty d\tilde{\omega}\Re[\tilde{\gamma}(\tilde{\omega}+i0^+)]\hbar\tilde{\omega}\sin[\tilde{\omega}(\tau -\tau^\prime)].
\end{eqnarray}
Using Eq.(\ref{green function}) and doing the integrals over $\tau$ and $\tau^\prime$, we obtain for the antisymmetric part
\begin{eqnarray}
\mathcal{A}(t,t^\prime) &=& \frac{\hbar}{i m\pi} \int_{-\infty} ^{+\infty} d\tilde{\omega}~\tilde{\omega}\Re[\frac{\tilde{\gamma}(\tilde{\omega})}{m}]\chi(\tilde{\omega})\chi^* (\tilde{\omega})e^{-i\tilde{\omega}(t-t^\prime)} \nonumber\\
&=& -\frac{\hbar}{2 m\pi} \int_{-\infty} ^{+\infty} d\tilde{\omega}[\chi(\tilde{\omega}) - \chi^* (\tilde{\omega})]e^{-i\tilde{\omega}(t-t^\prime)} .
\label{anti-response}
\end{eqnarray}
Unlike the symmetrized part, the integral in Eq.(\ref{anti-response}) is completely devoid of the temperature dependent $\coth(\beta\hbar\tilde{\omega}/2)$ term. After doing a complex contour integration we obtain
\begin{equation}
\mathcal{A}(t,t^\prime)= -\frac{\hbar}{m}\bigg\{\frac{(\omega_D -\lambda_1)e^{-\lambda_1 (t-t^\prime)}}{(\lambda_1 -\lambda_2)(\lambda_1 -\lambda_3)} + \frac{(\omega_D -\lambda_2)e^{-\lambda_2 (t-t^\prime)}}{(\lambda_2 -\lambda_3)(\lambda_2 -\lambda_1)}  + \frac{(\omega_D -\lambda_3)e^{-\lambda_3 (t-t^\prime)}}{(\lambda_3 -\lambda_1)(\lambda_3 -\lambda_2 )} \bigg\} .
\end{equation}
In the strict ohmic case ($\omega_D \rightarrow \infty$), the antisymmetric part becomes
\begin{equation}
\mathcal{A}(t,t^\prime)=\frac{\hbar}{m(\lambda_1 -\lambda_2)}\bigg\{e^{-\lambda_1 (t-t^\prime)} - e^{-\lambda_2 (t-t^\prime)}  \bigg\} ,
\label{antisymmetric-final}
\end{equation}
which is simplified further to yield
\begin{equation}
\mathcal{A}(t)= -\frac{\hbar}{m\Omega}e^{-\frac{\bar{\gamma} }{2}|t|}\sin\Omega t ,
\label{anti-anti}
\end{equation}
where $\Omega = \sqrt{\omega_0 ^2 -\frac{\bar{\gamma}^2}{4}}$. This result matches with the corresponding result for a quantum damped harmonic oscillator with $\gamma$ in the damped oscillator result being replaced by $\bar{\gamma}$ (which contains the magnetic contribution). For $\bar{\gamma} \gg 2\omega_0$, we write $\Omega \sim i\bar{\gamma}/2 - i\omega_0 ^2 /\bar{\gamma}$, so that $\mathcal{A}(t) \sim -(\hbar / m\bar{\gamma})e^{-\omega_0 ^2 t / \bar{\gamma}}$. It is very much clear from Eq.(\ref{antisymmetric-final}) that for equal times $t^\prime = t$, the antisymmetric part vanishes. Moreover the antisymmetric part is related to the commutator and it should be zero in the classical limit. This is obvious in the limit of $\hbar \rightarrow 0$ in Eq.(\ref{anti-anti}). It should be also noted that the antisymmetric part does not contain specific quantum effects (albeit an $\hbar$ factor appearing in it), but it is directly related to the classical response function (cf., Eq.(\ref{anti-response})) by the formula
\begin{equation}
\mathcal{A}(t,t^{\prime}) = -\frac{\hbar}{m}\{\chi(t-t^{\prime})-\chi^* (t-t^{\prime}) \} = -\frac{\hbar}{m}\{\chi(t-t^{\prime})-\chi (t^{\prime}-t) \},
\label{commutator-response-odd}
\end{equation}
and this implies that
\begin{equation}
\langle [z(t),z^\dagger (t^{\prime})] \rangle = \frac{2\hbar}{im}\{\chi(t-t^{\prime})-\chi (t^{\prime}-t) \},
\end{equation}
which is in connection with the linear response theory and the fluctuation-dissipation theorem. Using the property of the response function $\chi_{\rm odd} (t) = 1/2 [\chi(t)-\chi(-t)]$ and $\chi(t)=2\Theta(t)\chi_{\rm odd} (t)$, from Eq.(\ref{commutator-response-odd}), we write 
\begin{equation}
\chi(t) = -\frac{m}{\hbar}\Theta(t)\mathcal{A}(t) ,
\end{equation}
where $\Theta(t)$ is the Heaviside theta function. The immediate conclusion one can make out of the above relation is that the retarded Green's function or the response function is very closely related to the commutator of the position coordinates of the system. The linearity of the system guarantees that the commutators appearing in the problem are basically c-numbers. This emphasizes the fact that the retarded Green's functions are temperature independent. We have seen the advantage of having the retarded Green's function for calculations based on the  equations of motion for the operator of our interest. Nevertheless this retarded Green's function has to be distinguished from the time-ordered Green's functions, the latter being usually used in statistical physics and in the development of diagrammatic perturbation expansions.
\subsubsection{Fluctuation-dissipation theorem}
We now study the fluctuation-dissipation relation for the charged harmonic oscillator in the presence of a uniform and homogeneous magnetic field and a quantum heat bath consisting of non-interacting harmonic oscillators. The fluctuation-dissipation theorem (FDT) provides us with a relation between the linear response of the system to an external force or perturbation and the fluctuations in equilibrium. We can write the dynamical susceptibility $\chi(\omega)$ into its real and imaginary parts $\chi^{\prime} (\omega)$ and  $\chi^{\prime\prime} (\omega)$ respectively as follows
\begin{equation}
\chi(\omega) = \chi^{\prime} (\omega) + i \chi^{\prime\prime} (\omega) ,
\end{equation}
where the real and imaginary parts of $\chi(\omega)$ can be obtained from Eq.(\ref{laplace}) and are given by
\begin{equation}
\chi^{\prime} (\omega) = \frac{\omega_0 ^2 - \omega^2 + \omega\omega_c}{(\omega_0 ^2 - \omega^2 + \omega\omega_c)^2 + \omega^2 \frac{\tilde{\gamma}^2 (\omega)}{m^2}},~~{\rm and}~~\chi^{\prime\prime} (\omega) = \frac{\omega\tilde{\gamma}(\omega)/m}{(\omega_0 ^2 - \omega^2 + \omega\omega_c)^2 + \omega^2 \frac{\tilde{\gamma}^2 (\omega)}{m^2}}.
\end{equation}
Making use of the complex conjugate $\chi^* (\omega)$, we write
\begin{eqnarray*}
\chi^{\prime} (\tilde{\omega}) &=& \frac{1}{2}[\chi(\tilde{\omega})+\chi^* (\tilde{\omega})]=\frac{1}{2}[\chi(\tilde{\omega})+\chi(-\tilde{\omega})] , \\
\chi^{\prime \prime} (\tilde{\omega}) &=& \frac{1}{2i}[\chi(\tilde{\omega})-\chi^* (\tilde{\omega})]=\frac{1}{2i}[\chi(\tilde{\omega})-\chi(-\tilde{\omega})] ,
\end{eqnarray*}
so that
\begin{equation}
\chi(\tilde{\omega}) - \chi^* (\tilde{\omega}) = 2 i \chi^{\prime \prime} (\tilde{\omega}).
\label{short-relation}
\end{equation}
Using Eq.(\ref{short-relation}) we write (for $t^{\prime} =0$) Eq.(\ref{symmetric-fdt}) as
\begin{equation}
\mathcal{S}(t) = \frac{\hbar}{m \pi}\int_{-\infty}^{+\infty}
d\tilde{\omega}~\chi^{\prime \prime} (\tilde{\omega}) \coth \bigg(\frac{\beta\hbar\tilde{\omega}}{2} \bigg) e^{-i\tilde{\omega} t} ,
\label{symmetric-fdt1}
\end{equation}
which immediately yields
\begin{equation}
\mathcal{S}(\tilde{\omega}) = \frac{2\hbar}{m} \coth \bigg(\frac{\beta\hbar\tilde{\omega}}{2} \bigg) \chi^{\prime \prime} (\tilde{\omega}).
\end{equation}
This quantum mechanical relation has two limiting situations to study. In the extreme quantum limit where $k_B T \ll \hbar\tilde{\omega}$, we have the pure quantum fluctuations
\begin{equation}
\mathcal{S}(\tilde{\omega}) = \frac{2\hbar}{m} \chi^{\prime \prime} (\tilde{\omega}).
\end{equation}
The non-analyticity of the function $\mathcal{S}(\tilde{\omega})$ at the origin implies that the position autocorrelation function decays algebraically with time at $T=0$ as we have seen in Eq.(\ref{algebraic}). For a harmonic oscillator heat bath the imaginary part of the dynamical susceptibility $\chi^{\prime \prime} (\tilde{\omega})$ is a temperature independent quantity. For low frequencies or high temperatures where $k_B T \gg \hbar\tilde{\omega}$ we have
\begin{equation}
\mathcal{S}(\tilde{\omega}) = \frac{4}{m\beta} \frac{\chi^{\prime \prime} (\tilde{\omega})}{\tilde{\omega}},
\label{high-temp-ds}
\end{equation}
which is the pure classical result. In the absence of the magnetic field ($\omega_c =0$) we have from Eq.(\ref{high-temp-ds})
\begin{equation}
\mathcal{S}(\tilde{\omega}) = \frac{4}{m\beta} \frac{\tilde{\gamma}(\tilde{\omega})}{(\omega_0 ^2 - \tilde{\omega}^2)^2 + \tilde{\omega}^2 \tilde{\gamma}^2 (\tilde{\omega})} ,
\end{equation}
which is the fluctuation spectrum of the damped harmonic oscillator in thermal equilibrium. Similarly for the antisymmetric part of the position auto-correlation function, we write (cf., Eq.(\ref{anti-response}))
\begin{equation}
\mathcal{A}(\tilde{\omega}) = -\frac{2i\hbar}{m}\chi^{\prime\prime} (\tilde{\omega}).
\end{equation}
Hence we have the relation
\begin{equation}
\mathcal{C} (\tilde{\omega}) = \mathcal{S}(\tilde{\omega})+i\mathcal{A}(\tilde{\omega})= \frac{2\hbar}{m}\chi^{\prime\prime}(\tilde{\omega}) \bigg\{\coth\bigg(\frac{\beta\hbar\tilde{\omega}}{2} \bigg) +1\bigg\} .
\label{fdt-fourier-space}
\end{equation}
By using the definition 
\begin{equation}
\frac{1}{1-e^{-\beta\hbar\tilde{\omega}}} = \frac{1}{2}+\frac{1}{2}\coth\bigg(\frac{\beta\hbar\tilde{\omega}}{2} \bigg) ,
\end{equation}
we write
\begin{equation}
\mathcal{C} (\tilde{\omega}) = \frac{4\hbar}{m}\frac{\chi^{\prime\prime}(\tilde{\omega})}{1-e^{-\beta\hbar\tilde{\omega}}},
\end{equation}
which is the \textit{fluctuation-dissipation} theorem in the context of \textit{dissipative Landau diamagnetism}. Here the position autocorrelation function $\mathcal{C} (\tilde{\omega})$ is attributed to the spontaneous fluctuations of our system, while $\chi^{\prime\prime}(\tilde{\omega})$ determines the energy dissipation in the system due to work done by an external weak force.

Now we write the correlation function $\mathcal{C} (t,t^{\prime})$ using Eq.(\ref{fdt-fourier-space}) as
\begin{equation}
\mathcal{C} (t,t^{\prime}) = \frac{\hbar}{m\pi}\int_{-\infty}^\infty d\tilde{\omega}~\chi^{\prime \prime} (\tilde{\omega}) \bigg\{\coth \bigg(\frac{\beta\hbar\tilde{\omega}}{2} \bigg)+1\bigg\} e^{-i\tilde{\omega} (t-t^{\prime})}.
\label{correlation-stationarity}
\end{equation}
Note that $\mathcal{C} (t,t^{\prime})$ is a complex quantity and does not have simple classical analogue. But $\chi^{\prime} (\omega)$ and $\chi^{\prime\prime} (\omega)$ can be interpreted classically. The real and the imaginary parts of the unequal time correlation function $\mathcal{C} (t,t^{\prime})$ can now be easily written according to Eq.(\ref{real and imaginary}) as
\begin{equation}
\langle x(t)x(t^{\prime}) + y(t)y(t^{\prime}) \rangle  = \frac{\hbar}{m\pi}\int_{-\infty}^\infty d\tilde{\omega}~\chi^{\prime \prime}(\tilde{\omega}) \bigg\{\coth \bigg(\frac{\beta\hbar\tilde{\omega}}{2} \bigg)+1\bigg\} \cos[\tilde{\omega} (t-t^{\prime})],
\label{equal}
\end{equation}
and
\begin{equation}
\langle x(t)y(t^{\prime}) - y(t)x(t^{\prime}) \rangle  = \frac{\hbar}{m\pi}\int_{-\infty}^\infty d\tilde{\omega}~\chi^{\prime\prime} (\tilde{\omega}) \bigg\{\coth \bigg(\frac{\beta\hbar\tilde{\omega}}{2} \bigg)+1\bigg\} \sin[\tilde{\omega} (t-t^{\prime})].
\label{unequal}
\end{equation}
A closer look at the Eqs.(\ref{correlation-stationarity}), (\ref{equal}) and (\ref{unequal}) tells us that the correlation function depends on the time difference $(t-t^{\prime})$ implying stationarity. Therefore $t^{\prime}$ can be set equal to zero, which readily gives
\begin{equation}
\langle x(t)x(0) + y(t)y(0) \rangle  = \frac{\hbar}{m\pi}\int_{-\infty}^\infty d\tilde{\omega}~\chi^{\prime \prime} (\tilde{\omega}) \bigg\{\coth \bigg(\frac{\beta\hbar\tilde{\omega}}{2} \bigg)+1\bigg\} \cos[\tilde{\omega} t],
\end{equation}
and
\begin{equation}
\langle x(t)y(0) - y(t)x(0) \rangle  = \frac{\hbar}{m\pi}\int_{-\infty}^\infty d\tilde{\omega}~\chi^{\prime \prime} (\tilde{\omega}) \bigg\{\coth \bigg(\frac{\beta\hbar\tilde{\omega}}{2} \bigg)+1\bigg\} \sin[\tilde{\omega} t].
\end{equation}
The equal time correlation function can be found out in the limit of $t^{\prime} = t$ in the above equations, and only Eq.(\ref{equal}) contributes and Eq.(\ref{unequal}) vanishes as we have discussed in Eq.(\ref{real and imaginary}). The equilibrium position dispersion can be obtained by setting $t^{\prime} = t$ in Eq.(\ref{correlation-stationarity}) and is given by (with the knowledge that the anti-symmetrized part does not contribute)
\begin{equation}
\mathcal{C}  = \langle {\bf{r}}^2 \rangle = \langle x^2 + y^2 \rangle = \frac{\hbar}{m\pi}\int_{-\infty}^\infty d\tilde{\omega}~\chi^{\prime \prime} (\tilde{\omega}) \coth \bigg(\frac{\beta\hbar\tilde{\omega}}{2} \bigg).
\end{equation}
We can also write the equilibrium position dispersion using the relation $N(\omega)+1/2 = 1/2 \coth \bigg(\frac{\beta\hbar\tilde{\omega}}{2} \bigg)$, where $N(\omega)=1/[e^{-\beta\hbar\tilde{\omega}}-1]$, as
\begin{equation}
\mathcal{C}  = \frac{2\hbar}{m\pi}\int_{-\infty}^\infty d\tilde{\omega}~\chi^{\prime \prime} (\tilde{\omega}) [N(\tilde{\omega})+\frac{1}{2}] ,
\end{equation}
which is now the sum of the thermal contribution and the vacuum contribution. This infinite influence of the zero point fluctuations of the heat bath is a characteristic of all quantum noise problems.

We have seen that the dynamical susceptibility $\chi(\omega)$ is purely a classical quantity and the mean values obey the Ehrenfest's theorem. Quantum mechanics is entering into the autocorrelation functions only and solely by the fluctuation-dissipation theorem. Typically for a stationary stochastic process, the time correlations for the dynamical variables actually do not depend on absolute times. The time translational invariance is always preserved and is given by the property $\langle z(t)z(t^{\prime}) \rangle = \langle z(t-t^{\prime})z(0) \rangle$. Also we see that the response function and the equilibrium autocorrelation function of the position are related by the fluctuation-dissipation theorem. Since our system is a linear system, the underlying stochastic process is a stationary Gaussian process. Because for linear systems, the response to an external perturbation is actually linear in nature for arbitrary strength of the perturbation. With these points we justified the ``stochastic modeling" for the dissipative charged harmonic oscillator in a magnetic field.
\subsubsection{Equilibrium position dispersion}
Here we discuss the position autocorrelation function or dispersion at equilibrium. For $t^{\prime}=t$, we have seen that the time dependence disappeared completely from the position autocorrelation function and, we obtain, after a contour integration, the equilibrium value of the position autocorrelation as\cite{jishad2}
\begin{equation}
\label{zeta-equilibrium}
\langle {\bf{r}}^2 \rangle = \langle zz^{\dagger} \rangle = \frac{2}{m\beta\omega_0 ^2}+\frac{\hbar}{m\pi}\sum_{j=1}^3 \bigg[q_j \psi \bigg(1+ \frac{\lambda_j}{\nu} \bigg) + q^{\prime} _j \psi \bigg(1+ \frac{\lambda^{\prime} _j}{\nu} \bigg) \bigg],
\end{equation}
where $\psi(1+z)$ is the digamma function, $\nu =\nu_1 = \frac{2\pi}{\beta\hbar}$, and 
\begin{equation}
\label{first-check}
 q_1 = \frac{(\lambda_1 -\omega_D)}{(\lambda_1 -\lambda_2)(\lambda_1 -\lambda_3)},
\end{equation}
$q_2$ and $q_3$ are obtained by a cyclic permutation of roots in the above equation. The quantity $q^{\prime}_j$ is obtained by priming the $\lambda_j$s in Eq.(\ref{first-check}). In the absence of magnetic field ($\omega_c  =0$) as well as dissipation ($\gamma =0$), Eq.(\ref{zeta-equilibrium}) reduces to $\langle {\bf{r}}^2 \rangle  = \hbar /m\omega_0 \coth(\beta\hbar\omega_0 / 2)$, which is the familiar expression in the undamped case. It is easy to write down the mean squared value of the position in terms of the Matsubara frequencies $\nu_n$ using the definition of the digamma functions. Hence $\langle {\bf{r}}^2 \rangle$ can be written as
\begin{equation}
\langle {\bf{r}}^2 \rangle = \langle {\bf{r}}^2 \rangle _{classical} +\langle {\bf{r}}^2 \rangle _{quantum} =  \frac{2}{m\beta\omega_0 ^2}+ \frac{4}{m\beta}\sum_{n=1}^\infty \frac{(\nu_n ^2 + \omega_0 ^2 + \frac{\nu_n \gamma\omega_D}{(\nu_n +\omega_D)})}{(\nu_n ^2 + \omega_0 ^2 + \frac{\nu_n \gamma\omega_D}{(\nu_n +\omega_D)})^2 + \omega_c ^2 \nu_n ^2}.
\end{equation}
Using the fact that $\gamma(\nu_n) > 0$ and $(\nu_n ^2 + \omega_0 ^2 + \frac{\nu_n \gamma\omega_D}{(\nu_n +\omega_D)}) > 0~(n=0, 1, 2, .....)$, $\langle {\bf{r}}^2  \rangle$ decreases monotonically with the increasing value of the strength of the magnetic field, i.e.,
\begin{equation}
\frac{\partial}{\partial B} \langle {\bf{r}}^2  \rangle < 0.
\end{equation}
This implies the fact that the dissipative charged oscillator in a magnetic field is still generally diamagnetic and is unaltered by the presence of an arbitrary heat bath. It has been proved that, for strict ohmic dissipative heat bath, when the magnetic field is stronger than a certain critical value, weak dissipation actually delocalizes the oscillation of the charged  particle (unlike the problem without the magnetic field where dissipation always leads to enhanced localization)\cite{li3}. Also, strong dissipation localizes the motion and the magnetic field just enhances the localization. Now we discuss the low and high temperature behavior of the mean squared position. At low temperatures the mean squared position is given by
\begin{equation}
\langle {\bf{r}}^2 \rangle = \frac{2\pi}{3}\frac{\hbar\gamma}{m\omega_0 ^2} \bigg(\frac{1}{\beta\hbar\omega_0}\bigg)^2 + \frac{\hbar}{m\pi}\sum_{j=1}^3 [q_j \ln \lambda_j + q^{\prime} _j \ln \lambda^{\prime} _j],
\label{mean-square-pos1}
\end{equation}
so that
\begin{equation}
\langle {\bf{r}}^2 \rangle _{T=0} = \frac{\hbar}{m\pi}\sum_{j=1}^3 [q_j \ln \lambda_j + q^{\prime} _j \ln \lambda^{\prime} _j].
\end{equation}

In the limit of $\omega_D \rightarrow \infty$, the low-temperature behavior of $\langle {\bf{r}}^2 \rangle$ can be obtained from Eq.(\ref{mean-square-pos1}) as
\begin{equation}
\langle {\bf{r}}^2 \rangle = \frac{2\pi}{3}\frac{\hbar\gamma}{m\omega_0 ^2} \bigg(\frac{1}{\beta\hbar\omega_0}\bigg)^2 + \frac{\hbar}{m\pi}\bigg[\frac{1}{\Lambda}\ln \bigg(\frac{\lambda_1}{\lambda_2}\bigg) + \frac{1}{\Lambda^{\prime}}\ln \bigg(\frac{\lambda^{\prime} _1}{\lambda^{\prime} _2}\bigg)\bigg],
\label{position-expanded}
\end{equation}
where $\Lambda = \sqrt{\bar{\gamma}^2-4\omega_0 ^2}$ and $\Lambda^{\prime} =  \sqrt{\bar{\gamma}^{*2}-4\omega_0 ^2}$. The Eq.(\ref{position-expanded}) correctly reproduces $\langle {\bf{r}}^2 \rangle _{T=0} = \hbar / m\omega_0$ for the dispersion in position coordinate of an undamped oscillator in its ground state. For the strong damping, i.e., $\bar{\gamma} \gg 2\omega_0$, we obtain
\begin{equation}
\langle {\bf{r}}^2 \rangle = \frac{2\pi}{3}\frac{\hbar\gamma}{m\omega_0 ^2} \bigg(\frac{1}{\beta\hbar\omega_0}\bigg)^2 + \frac{2\hbar}{m\pi}\bigg[\frac{1}{\bar{\gamma}}\ln \bigg(\frac{\bar{\gamma}}{\omega_0}\bigg) + \frac{1}{\bar{\gamma}^{*}}\ln \bigg(\frac{\bar{\gamma}^{*}}{\omega_0}\bigg)\bigg] + \mathcal{O}(T^4),
\end{equation}
which in the absence of magnetic field ($\omega_c =0$) yields a form
\begin{equation}
\langle {\bf{r}}^2 \rangle = \frac{4\hbar}{m\pi\gamma} \ln \bigg(\frac{\gamma}{\omega_0}\bigg) + \frac{2\pi}{3}\frac{\hbar\gamma}{m\omega_0 ^2} \bigg(\frac{1}{\beta\hbar\omega_0}\bigg)^2 + \mathcal{O}(T^4),
\end{equation}
which is the corresponding result for a damped harmonic oscillator in two dimensions. Since $\langle {\bf{r}}^2 \rangle = \langle x^2  \rangle + \langle y^2 \rangle$ and $\langle x^2  \rangle = \langle y^2 \rangle$, we write in general
\begin{equation}
\langle x^2 \rangle = \frac{1}{m\beta\omega_0 ^2} + \frac{2}{m\beta}\sum_{n=1}^\infty \frac{(\nu_n ^2 + \omega_0 ^2 + \frac{\nu_n \gamma\omega_D}{(\nu_n +\omega_D)})}{(\nu_n ^2 + \omega_0 ^2 + \frac{\nu_n \gamma\omega_D}{(\nu_n +\omega_D)})^2 + \omega_c ^2 \nu_n ^2}.
\end{equation}
Again in the limit of $\omega_D \rightarrow \infty$, $\langle x^2 \rangle$ at low temperatures behaves as
\begin{equation}
\langle x^2 \rangle = \frac{\pi}{3}\frac{\hbar\gamma}{m\omega_0 ^2} \bigg(\frac{1}{\beta\hbar\omega_0}\bigg)^2 + \frac{2\hbar}{\pi m a }\bigg[ \sqrt{\frac{a-b}{2}}{\rm \tan}^{-1} \bigg( \frac{1}{\gamma}\sqrt{\frac{a-b}{2}} \bigg) + \frac{1}{2}\sqrt{\frac{a+b}{2}}\ln \bigg(\frac{\gamma /2 + 1/2 \sqrt{(a+b)/2}}{\gamma /2 - 1/2 \sqrt{(a+b)/2}}\bigg)\bigg],
\label{lie-formula}
\end{equation}
which is consistent with the result obtained by Li et al\cite{li3}. Here we define $a=\sqrt{(\gamma^2 - \omega_c ^2 -4\omega_0 ^2)^2 + 4\gamma^2 \omega_c ^2}$ and $b=(\gamma^2 - \omega_c ^2 -4\omega_0 ^2)$. Switching the magnetic field off ($\omega_c = 0$) produces the respective result for a damped quantum harmonic oscillator\cite{weissbook}. The leading correction term in Eq.(\ref{lie-formula}) is proportional to $T^2$ because of the ohmic nature of the heat bath. Also note that this correction term is independent of the magnetic field.
\begin{figure}
  \centering
  \subfloat[]{\label{fig:low-c-q}\includegraphics[width=0.45\textwidth]{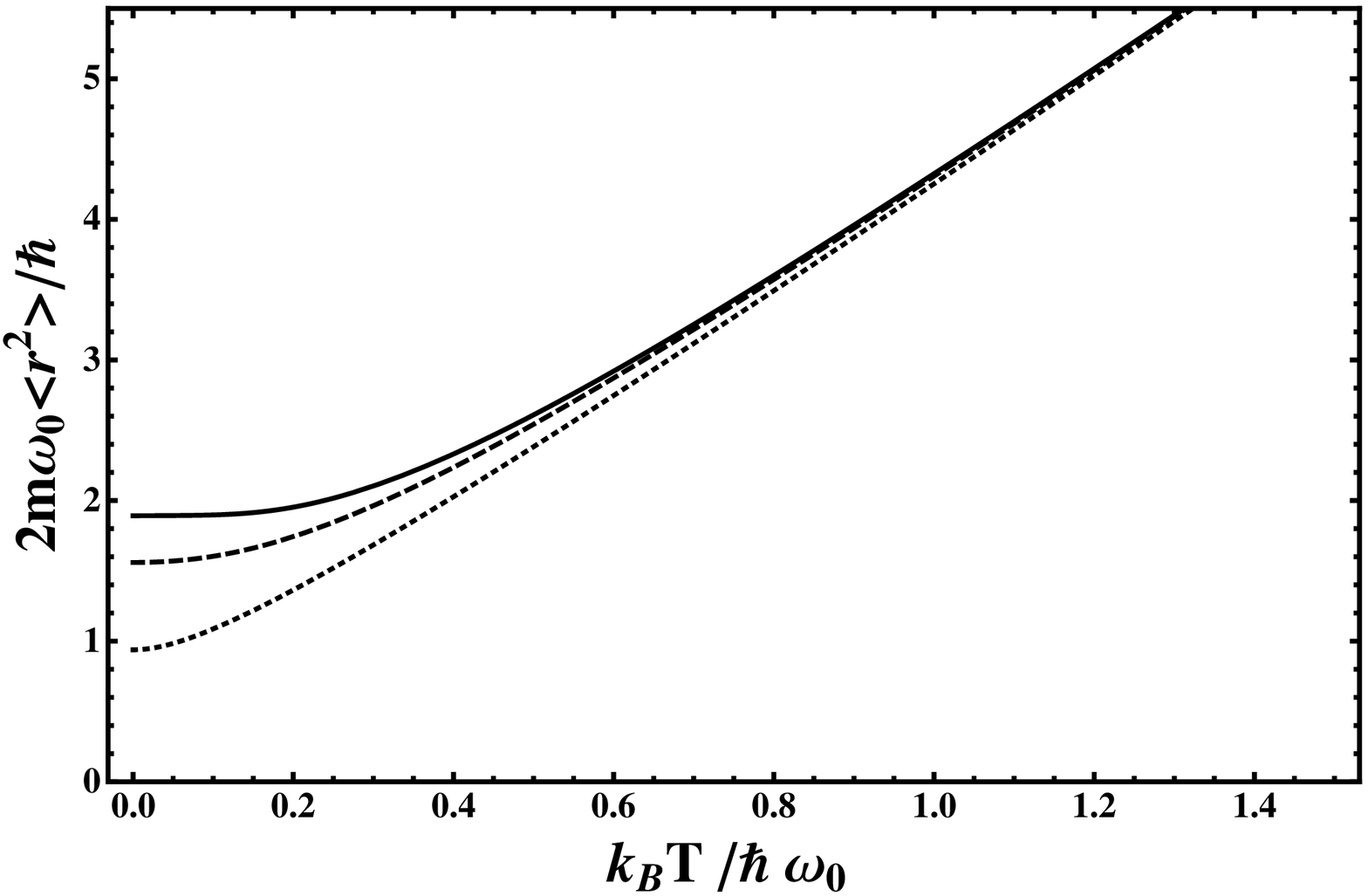}} \hspace {0.2cm}               
  \subfloat[]{\label{fig:high-c-q}\includegraphics[width=0.45\textwidth]{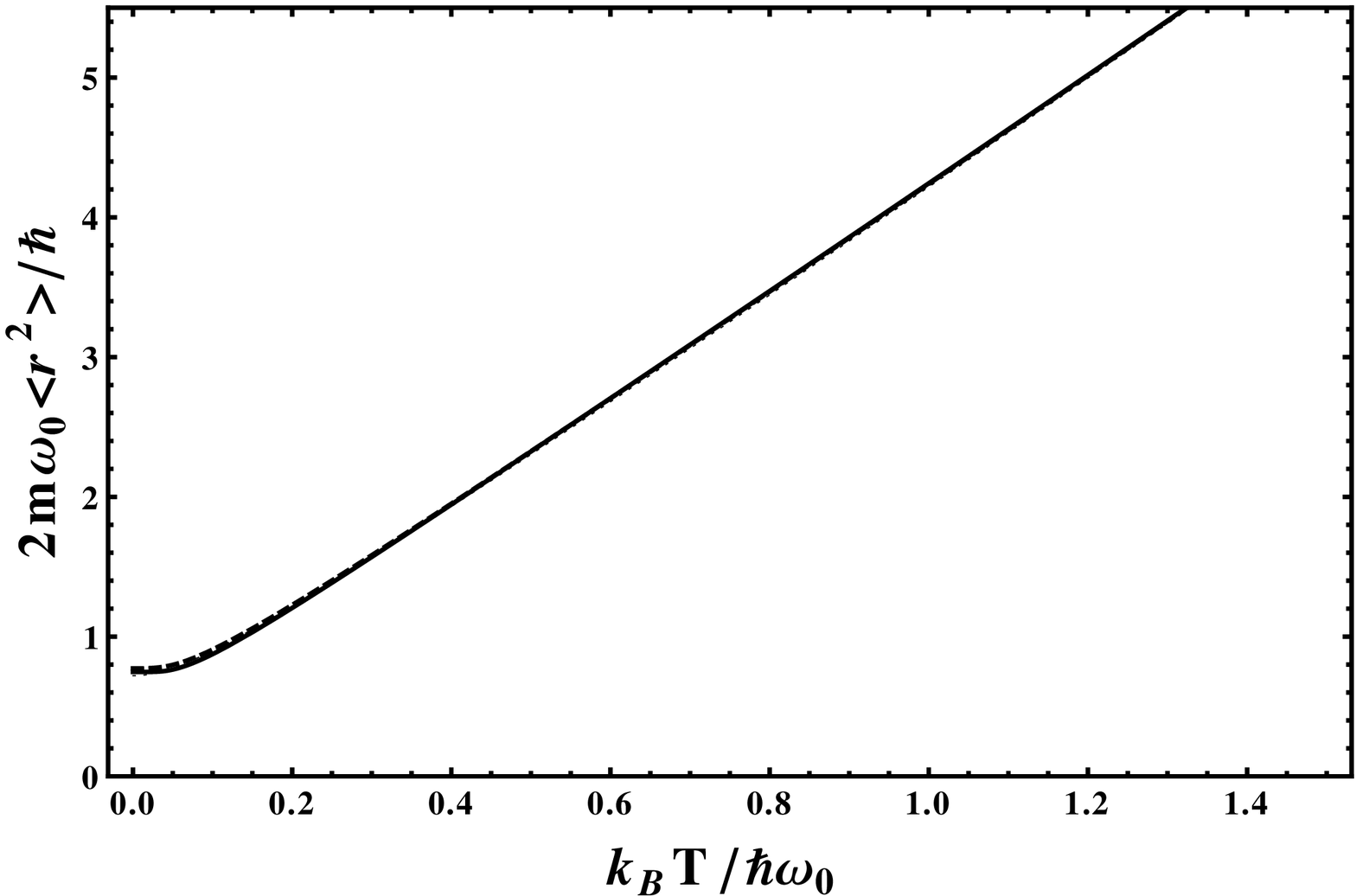}}
  \caption{The normalized equilibrium dispersion of position with respect to the dimensionless temperature $k_B T / \hbar\omega_0$ for (a) low magnetic field ($\omega_c / \omega_0 = 0.5$) and (b) for high magnetic field ($\omega_c / \omega_0 = 5.0$) is given above. The thick curve is for $\gamma / \omega_0 = 0.1$, the dashed one is for $\gamma / \omega_0 = 1.0$ and dotted one is for $\gamma / \omega_0 = 5.0$. Here we set $\omega_D / \omega_0 = 10.0$.}
  \label{avg-pos}
\end{figure}

Meanwhile at high temperatures, the mean squared position varies according to
\begin{equation}
\langle {\bf{r}}^2 \rangle = \frac{2}{m\beta\omega_0 ^2}\bigg[1+\frac{1}{24}(\beta\hbar\omega_0)^2 + \mathcal{O}\bigg(\frac{1}{T}\bigg)^3 \bigg].
\end{equation}
Also
\begin{equation}
\langle x^2 \rangle = \frac{1}{m\beta\omega_0 ^2}+ \mathcal{O}\bigg(\frac{1}{T}\bigg).
\end{equation}
It is to be noted here that, upto the first order, the high temperature value of the position dispersion is independent of the dissipation strength $\gamma$, the Drude cutoff $\omega_D$ and the magnetic field $
\omega_c$.  Moreover the mean squared position restores the classical equipartition result at very high temperatures.
\subsubsection{Classical diffusive behavior of the free quantum charged particle in a magnetic field}
Having discussed about the dissipative charged oscillator in a magnetic field, we are here to check whether in the limit of $\omega_0 \rightarrow 0$ one can see the classical long time diffusive behavior of the free charged quantum particle under the influence of a heat bath. It is well understood from Eq.(\ref{zeta-equilibrium}) for the equilibrium position dispersion that the removal of the oscillator potential produces a singularity. Moreover from the symmetric part of the position autocorrelation function given in Eq.(\ref{symmetric-fdt}), one can take out the confinement and obtain the long time behavior. But as $\omega_0 \rightarrow 0$, one of the roots each from each set, say, $\lambda_1$ and $\lambda_1 ^{\prime}$ are proportional to $\omega_0$. Therefore, the exponentials containing $\lambda_1$ and $\lambda_1 ^{\prime}$ have to be treated carefully. But we obtained Eq.(\ref{symmetric-fdt}) by assuming that except the time difference $t-t^{\prime}$, $t$ and $t^{\prime}$ are very large and subsequently all the exponential terms except $e^{-i\tilde{\omega}(t-t^{\prime})}$ vanish. Therefore the diffusive result one may obtain from Eq.(\ref{symmetric-fdt}) represents a part of the original result. Hence in this case, in order to see the long time diffusive behavior we need to resort to Eq.(\ref{zeta-trignometric}). Here we work at equal times ($t^{\prime} =t$). Hence in addition to the steady (time-independent) part as in Eq.(\ref{zeta-equilibrium}) we have contributions from another twenty seven integrals. But for our present purpose few of them are relevant and the all the results obtained after evaluating the rest of the integrals vanish in the limit of $t\rightarrow\infty$. So we do not account for them here. What is more important for us is to look at those terms which contain $\lambda_1$ and $\lambda^{\prime} _1$. We write
\begin{eqnarray}
\langle x^2 (t) + y^2 (t) \rangle &=& \frac{2}{m\beta\omega_0 ^2} + \frac{2}{m\beta}\sum_{n=1}^\infty \bigg\{  \frac{(\nu_n +\omega_D)}{(\nu_n + \lambda_1)(\nu_n + \lambda_2)(\nu_n + \lambda_3)} + \frac{(\nu_n +\omega_D)}{(\nu_n + \lambda^{\prime} _1)(\nu_n + \lambda^{\prime} _2)(\nu_n + \lambda^{\prime} _3)}  \bigg\}\nonumber\\
&-& \frac{\mathcal{Q}}{m\pi \mathcal{M}\mathcal{M}^{\prime}} \int_{-\infty} ^{+\infty} d\tilde{\omega}\frac{\gamma\omega_D ^2}{\omega_D ^2 + \tilde{\omega}^2}\hbar\tilde{\omega}\coth\bigg(\frac{\beta\hbar\tilde{\omega}}{2}\bigg)\bigg\{\frac{e^{-(i\tilde{\omega}+\lambda^{\prime} _1)t}}{(\tilde{\omega}+i\lambda_1)(\tilde{\omega}-i\lambda^{\prime} _1)} \nonumber\\
&+& \frac{e^{(i\tilde{\omega}-\lambda_1)t}}{(\tilde{\omega}+i\lambda_1)(\tilde{\omega}-i\lambda^{\prime} _1)} -\frac{e^{-(\lambda_1+\lambda_1)t}}{(\tilde{\omega}+i\lambda_1)(\tilde{\omega}-i\lambda^{\prime} _1)}  \bigg\} ,
\end{eqnarray}
or
\begin{eqnarray}
\langle x^2 (t) + y^2 (t) \rangle &=& \frac{2}{m\beta\omega_0 ^2} + \frac{\hbar}{m\pi}\sum_{j=1}^3 \bigg\{ q_j \psi \bigg(1+\frac{\lambda_j}{\nu} \bigg) + q^{\prime} _j\psi \bigg(1+\frac{\lambda^{\prime}_j}{\nu} \bigg)\bigg\}\nonumber\\
&-& \frac{\mathcal{Q}}{m\pi \mathcal{M}\mathcal{M}^{\prime}} \int_{-\infty} ^{+\infty} d\tilde{\omega}\frac{\gamma\omega_D ^2}{\omega_D ^2 + \tilde{\omega}^2}\hbar\tilde{\omega}\coth\bigg(\frac{\beta\hbar\tilde{\omega}}{2}\bigg)\bigg\{\frac{e^{-(i\tilde{\omega}+\lambda^{\prime} _1)t}}{(\tilde{\omega}+i\lambda_1)(\tilde{\omega}-i\lambda^{\prime} _1)} \nonumber\\
&+& \frac{e^{(i\tilde{\omega}-\lambda_1)t}}{(\tilde{\omega}+i\lambda_1)(\tilde{\omega}-i\lambda^{\prime} _1)} -\frac{e^{-(\lambda_1+\lambda_1)t}}{(\tilde{\omega}+i\lambda_1)(\tilde{\omega}-i\lambda^{\prime} _1)}  \bigg\} ,
\end{eqnarray}
where $q_j$'s, $\mathcal{M}$ and $\mathcal{M}^{\prime}$ are defined earlier. Here we call $\mathcal{Q}=(\lambda_1 -\omega_D)(\lambda^{\prime} _1 -\omega_D)(\lambda_2 - \lambda_3)(\lambda^{\prime} _2  -\lambda^{\prime} _3)$. The strategy is to evaluate the integrals first and then take the limit of $\omega_0 \rightarrow 0$ to obtain the free particle result. Choosing the proper complex contours for the residue evaluation, we obtain
\begin{eqnarray}
\langle x^2 (t) + y^2 (t) \rangle &=& \frac{2}{m\beta\omega_0 ^2} + \frac{\hbar}{m\pi}\sum_{j=1}^3 \bigg\{ q_j \psi \bigg(1+\frac{\lambda_j}{\nu} \bigg) + q^{\prime} _j\psi \bigg(1+\frac{\lambda^{\prime}_j}{\nu} \bigg)\bigg\}\nonumber\\
&-& \frac{2\pi \mathcal{Q}}{m\pi \mathcal{M}\mathcal{M}^{\prime}}\bigg\{\frac{\gamma\omega_D ^2}{\omega_D ^2 -\lambda_1 ^2}\bigg[i\hbar\lambda_1 \coth\bigg(\frac{i\beta\hbar\lambda_1}{2}\bigg) \bigg]\frac{e^{-(\lambda_1 + \lambda^{\prime} _1)t}}{(\lambda_1 +\lambda^{\prime} _1)}\nonumber\\
&-& \frac{\gamma\omega_D}{2}\bigg[i\hbar\omega_D \coth\bigg(\frac{i\beta\hbar\omega_D}{2}\bigg) \bigg]\frac{e^{-(\omega_D + \lambda^{\prime} _1)t}}{(\omega_D - \lambda_1)(\omega_D + \lambda^{\prime} _1)}\nonumber\\
&+& \frac{2}{\beta}\sum_{n=1}^\infty \frac{\gamma\omega_D ^2}{\omega_D ^2 -\nu_n ^2}\frac{\nu_n e^{-(\nu_n +\lambda^{\prime} _1)t}}{(\nu_n -\lambda_1)(\nu_n +\lambda^{\prime} _1)}\nonumber\\
&-& \frac{\gamma\omega_D}{2}\bigg[i\hbar\omega_D \coth\bigg(\frac{i\beta\hbar\omega_D}{2}\bigg) \bigg]\frac{e^{-(\omega_D + \lambda_1)t}}{(\omega_D + \lambda_1)(\omega_D - \lambda^{\prime} _1)}\nonumber\\
&+& \frac{2}{\beta}\sum_{n=1}^\infty \frac{\gamma\omega_D ^2}{\omega_D ^2 -\nu_n ^2}\frac{\nu_n e^{-(\nu_n +\lambda_1)t}}{(\nu_n + \lambda_1)(\nu_n - \lambda^{\prime} _1)}\nonumber\\
&+& \frac{\gamma\omega_D}{2}\bigg[i\hbar\omega_D \coth\bigg(\frac{i\beta\hbar\omega_D}{2}\bigg) \bigg]\frac{e^{-(\lambda_1 + \lambda^{\prime} _1)t}}{(\omega_D + \lambda_1)(\omega_D - \lambda^{\prime} _1)}\nonumber\\
&-& \frac{2}{\beta}\sum_{n=1}^\infty \frac{\gamma\omega_D ^2}{\omega_D ^2 -\nu_n ^2}\frac{\nu_n e^{-(\lambda_1 + \lambda^{\prime} _1)t}}{(\nu_n + \lambda_1)(\nu_n - \lambda^{\prime} _1)} \bigg\}.
\label{additional-zeta}
\end{eqnarray}
Now we take the limit of $\omega_0 \rightarrow 0$. In the vanishing oscillator potential limit, from the vieta equations given by Eq.(\ref{vieta}) and Eq.(\ref{vieta1}), we obtain
\begin{equation}
\lambda_1 = \frac{\omega_0 ^2}{(\gamma+i\omega_c)},~~{\rm and}~~\lambda_1 ^{\prime} = \frac{\omega_0 ^2}{(\gamma-i\omega_c)} ,
\end{equation}
which gives $\lambda_1 + \lambda_2 = \frac{2\gamma\omega_0 ^2}{\gamma^2 + \omega_c ^2}$ . We expand the exponential as $\exp[-(\lambda_1 + \lambda^{\prime} _1)t] \approx 1-(\lambda_1 + \lambda^{\prime} _1)t$ (neglecting higher order terms). While simplifying we get a term $-2/m\beta\omega_0 ^2$ which cancels the first term in Eq.(\ref{additional-zeta}). It therefore helps us to obtain a result devoid of the singularity. Therefore in the limit of $\omega_0 \rightarrow 0$, the final expression for the mean squared value of the position is given by
\begin{equation}
\begin{split}
\langle x^2 (t) + y^2 (t) \rangle = \langle {\bf{r}}^2 (t) \rangle   &= \frac{4\gamma}{m\beta(\gamma^2 + \omega_c ^2)}t + \frac{2}{m\beta}\sum_{n=1}^\infty \bigg\{\frac{(\nu_n + \omega_D)}{\nu_n (\nu_n + \lambda_2)(\nu_n + \lambda_3)} + \frac{(\nu_n + \omega_D)}{\nu_n (\nu_n + \lambda^{\prime} _2)(\nu_n + \lambda^{\prime} _3)}\bigg\} \\
&- \frac{\hbar\gamma}{m(\gamma^2 + \omega_c ^2)}\cot\bigg(\frac{\beta\hbar\omega_D}{2}\bigg)[1-2e^{-\omega_D t}] + \frac{4}{m\beta}\frac{\gamma\omega_D ^2}{(\gamma^2 + \omega_c ^2)}\sum_{n=1}^\infty \frac{[1-2e^{-\nu_n t}]}{\nu_n (\omega_D ^2 - \nu_n ^2)} .
\end{split}
\end{equation}
After a small rearrangement of the second term, we write
\begin{equation}
\begin{split}
\langle x^2 (t) + y^2 (t) \rangle = \langle {\bf{r}}^2 (t) \rangle   &= \frac{4\gamma}{m\beta(\gamma^2 + \omega_c ^2)}t + \frac{4}{m\beta}\sum_{n=1}^\infty \frac{(\nu_n + \frac{\gamma\omega_D}{\nu_n + \omega_D})}{\nu_n [(\nu_n + \frac{\gamma\omega_D}{\nu_n + \omega_D})^2 + \omega_c ^2 ]} \\
&- \frac{\hbar\gamma}{m(\gamma^2 + \omega_c ^2)}\cot\bigg(\frac{\beta\hbar\omega_D}{2}\bigg)[1-2e^{-\omega_D t}] + \frac{4}{m\beta}\frac{\gamma\omega_D ^2}{(\gamma^2 + \omega_c ^2)}\sum_{n=1}^\infty \frac{[1-2e^{-\nu_n t}]}{\nu_n (\omega_D ^2 - \nu_n ^2)} .
\end{split}
\end{equation}
which in the classical ($\hbar =0$) limit and for $\gamma t \gg 1$ and $t\gg \hbar/k_B T$, we obtain (in the limit of $\omega_D \rightarrow \infty$)
\begin{equation}
\langle {\bf{r}}^2 (t) \rangle = \frac{4k_B T \gamma}{m(\gamma^2 + \omega_c ^2)}t,
\label{sdg-result}
\end{equation}
and that in the absence of the magnetic field, ie., $\omega_c = 0$ yields the expected diffusive result
\begin{equation}
\langle {\bf{r}}^2 (t) \rangle = \frac{4k_B T}{m\gamma}t ,
\end{equation}
of a free particle in two dimensions. The Eq.(\ref{sdg-result}) is matching with the one obtained in \cite{singh}. Note here that the mean squared displacement of the dissipative quantum charged particle is asymptotically proportional to the time $t$ with diffusion constant $D=4k_B T \gamma /[m(\gamma^2 + \omega_c ^2)]$ as the coefficient, exactly similar to the classical case. The diffusion coefficient vanishes at zero temperature. Precisely the growth of the mean squared displacement is slowing down as $T$ approaches zero. The mean squared displacement of charged particle in the presence of a magnetic field in the quantum case was discussed in \cite{revathi} also. Moreover, it was given in \cite{kumar} that a quantum particle may show a superdiffusive behavior as well. At zero temperature and for $\gamma t \gg 1$, it has been shown that the diffusive behavior of the quantum Brownian particle is logarithmic in nature \cite{vinay}, a sign of strong subdiffusive behavior in the quantum regime.

The assumption of thermally uncorrelated initial states of the system and the heat bath at time $t=0$ leads to large transient evolution as we discussed earlier. This initial jolts allow the charged Brownian particle in a magnetic field to absorb an arbitrary amount of energy from the high frequency modes of the quantum heat bath and subsequently it can travel an arbitrary distance within a finite time window. This leads to a divergent contribution of the mean squared displacement. However, once we assume the times $t$ and $t^{\prime}$ to be very large, all the transient terms vanish, $\langle {\bf{r}}^ 2 (t) \rangle$ becomes convergent. Thereafter the diffusive behavior of the particle is given exactly by the same formula as one sees in classical stochastic process. This energy absorption and the subsequent travel of the Brownian particle can be well understood by the linearly rising nature of the term $\tilde{\omega}\coth(\beta\hbar\tilde{\omega} /2)$ as $\tilde{\omega}$ gets large. This indeed implies that they are a result of the very large zero point energies available in the very high frequency oscillators of the quantum heat bath. One can in principle say that the initial transient behavior is an essential phenomenon to simply destroy the artificially constructed product initial state of the full system-plus-bath arrangement.
\subsection{Velocity autocorrelation function}
\label{sec:5}
In this subsection we deal with the velocity autocorrelation function which is calculated from the unequal time correlation function $\mathcal{D}(t,t^{\prime})=\langle \dot{z}(t)\dot{z}^{\dagger} (t^{\prime}) \rangle$. This autocorrelation function, at equal times, is used to evaluate the equilibrium kinematic momentum dispersion which we will discuss in the forthcoming subsection. Using the definition $z=x+iy$, we write
\begin{eqnarray}
\mathcal{D}(t,t^{\prime}) &=& \langle \dot{z}(t)\dot{z}^{\dagger} (t^{\prime}) \rangle \nonumber\\
&=& \langle [\dot{x}(t)\dot{x}(t^{\prime})+\dot{y}(t)\dot{y}(t^{\prime})]\rangle - i\langle[\dot{x}(t)\dot{y}(t^\prime)-\dot{y}(t)\dot{x}(t^\prime)]\rangle \nonumber\\
&=& \langle[v_x (t) v_x (t^{\prime}) + v_y (t) v_y (t^{\prime})]\rangle -i\langle[v_x (t) v_y (t^\prime) - v_y (t) v_x (t^\prime)]\rangle \nonumber\\
&=& \langle [v_x (t) v_x (t^{\prime}) + v_y (t) v_y (t^{\prime})]\rangle -i \langle({\bf{v}}(t)\times {\bf{v}}(t^{\prime}))_z \rangle .
\end{eqnarray}
We can still define the symmetric and the commutator (anti-symmetric) parts of the unequal time correlation function $\mathcal{D}(t,t^{\prime})$ (similar to what we defined for $\mathcal{C}(t,t^{\prime})$) using $\dot{z}(t)$ and $\dot{z}^{\dagger} (t^\prime)$. The unequal time correlation function $\mathcal{D}(t,t^{\prime}) = \langle \dot{z}(t)\dot{z}^{\dagger}(t^{\prime}) \rangle$ can be obtained by taking the derivative of Eq.(\ref{correlation-stationarity}) with respect to $t$ and $t^{\prime}$ respectively. Hence, with the help of Eqs. (\ref{symmetric-fdt}) and (\ref{anti-response}), one can in principle write the symmetric and the antisymmetric (commutator) structures by taking derivatives with respect to $t$ and $t^\prime$. However, we do not derive them explicitly and separately here as these results are not of much interest here. We obtain, after taking the derivatives of Eq.(\ref{correlation-stationarity}) with respect to $t$ and $t^\prime$
\begin{equation}
\mathcal{D}(t,t^{\prime}) = \frac{\hbar}{m\pi}\int_{-\infty}^\infty d\tilde{\omega}~\tilde{\omega}^2 ~\chi^{\prime\prime} (\tilde{\omega}) \bigg\{\coth \bigg(\frac{\beta\hbar\tilde{\omega}}{2} \bigg)+1\bigg\} e^{-i\tilde{\omega} (t-t^{\prime})}.
\label{correlation2-stationarity}
\end{equation}
Subsequently the real and imaginary parts are given by
\begin{equation}
\langle v_x (t) v_x (t^{\prime}) + v_y (t) v_y (t^{\prime}) \rangle  = \frac{\hbar}{m\pi}\int_{-\infty}^\infty d\tilde{\omega}~\tilde{\omega}^2 ~\chi^{\prime\prime} (\tilde{\omega}) \bigg\{\coth \bigg(\frac{\beta\hbar\tilde{\omega}}{2} \bigg)+1\bigg\} \cos[\tilde{\omega} (t-t^{\prime})],
\label{real2}
\end{equation}
and
\begin{equation}
\langle ({\bf{v}}(t)\times {\bf{v}}(t^{\prime}))_z \rangle  = \frac{\hbar}{m\pi}\int_{-\infty}^\infty d\tilde{\omega}~\tilde{\omega}^2 ~\chi^{\prime\prime} (\tilde{\omega}) \bigg\{\coth \bigg(\frac{\beta\hbar\tilde{\omega}}{2} \bigg)+1\bigg\} \sin[\tilde{\omega} (t-t^{\prime})].
\label{imaginary2}
\end{equation}
For $t^{\prime}=0$, we write
\begin{equation}
\langle v_x (t) v_x (0) + v_y (t) v_y (0) \rangle  = \frac{\hbar}{m\pi}\int_{-\infty}^\infty d\tilde{\omega}~\tilde{\omega}^2 ~\chi^{\prime\prime} (\tilde{\omega}) \bigg\{\coth \bigg(\frac{\beta\hbar\tilde{\omega}}{2} \bigg)+1\bigg\} \cos\tilde{\omega} t,
\label{real3}
\end{equation}
and
\begin{equation}
\langle ({\bf{v}}(t)\times {\bf{v}}(0))_z \rangle  = \frac{\hbar}{m\pi}\int_{-\infty}^\infty d\tilde{\omega}~\tilde{\omega}^2 ~\chi^{\prime\prime} (\tilde{\omega}) \bigg\{\coth \bigg(\frac{\beta\hbar\tilde{\omega}}{2} \bigg)+1\bigg\} \sin\tilde{\omega} t.
\label{imaginary3}
\end{equation}
After doing a complex contour integration we obtain
\begin{equation}
\langle v_x (t) v_x (0) + v_y (t) v_y (0) \rangle  = -\frac{2}{m\beta}\bigg\{\frac{\lambda_1 (\omega_D - \lambda_1)}{(\lambda_1 - \lambda_2)(\lambda_1 - \lambda_3)}e^{-\lambda_1 t} + \frac{\lambda_2 (\omega_D - \lambda_2)}{(\lambda_2 - \lambda_1)(\lambda_2 - \lambda_3)}e^{-\lambda_2 t} + \frac{\lambda_3 (\omega_D - \lambda_3)}{(\lambda_3 - \lambda_2)(\lambda_3 - \lambda_1)}e^{-\lambda_3 t} \bigg\},
\label{real4}
\end{equation}
which for the strict ohmic damping ($\omega_D \rightarrow \infty$) limit yields
\begin{equation}
\langle v_x (t) v_x (0) + v_y (t) v_y (0) \rangle  = \frac{2}{m\beta\sqrt{\bar{\gamma}^2 -4\omega_0 ^2}} \{\lambda_1 e^{-\lambda_1 t} - \lambda_2 e^{-\lambda_2 t} \}.
\end{equation}
In the free particle limit ($\omega_0 = 0$) this becomes
\begin{equation}
\langle v_x (t) v_x (0) + v_y (t) v_y (0) \rangle  = \frac{2}{m\beta} e^{-\gamma |t|}\cos(\omega_c t).
\label{sdg-correlation1}
\end{equation}
Similarly we can calculate the average $\langle ({\bf{v}}(t)\times {\bf{v}}(0))_z \rangle$ and in the free particle limit this is given by
\begin{equation}
\langle ({\bf{v}}(t)\times {\bf{v}}(0))_z \rangle = \frac{2}{m\beta} e^{-\gamma |t|}\sin(\omega_c t).
\label{sdg-correlation2}
\end{equation}
Eqs. (\ref{sdg-correlation1}) and (\ref{sdg-correlation2}) are in good agreement with the result obtained by Dattagupta and Singh \cite{singh}. In the absence of the magnetic field ($\omega_c = 0$), we obtain (from Eq.(\ref{sdg-correlation1})) $\langle v_x (t) v_x (0) + v_y (t) v_y (0) \rangle  = \frac{2}{m\beta} e^{-\gamma |t|}$, which is in complete agreement with the velocity autocorrelation function for the free damped particle. At equal times ($t^{\prime}=t$), again we see that the imaginary part (Eq.(\ref{imaginary2})) vanishes making no contribution to the correlation and, only the real part (Eq.(\ref{real2})) contributes. From Eq.(\ref{correlation2-stationarity}), we write
\begin{equation}
\langle v_x ^2 + v_y ^2 \rangle  = \frac{\hbar}{m\pi}\int_{-\infty}^\infty d\tilde{\omega}~\tilde{\omega}^2 ~\chi^{\prime\prime} (\tilde{\omega}) \coth \bigg(\frac{\beta\hbar\tilde{\omega}}{2} \bigg).
\label{velocity-correlation-function}
\end{equation}
For strict ohmic damping the integral in Eq.(\ref{velocity-correlation-function}) diverges. This shows that the assumption of a memoryless heat bath is unphysical. Real physical systems exhibit a microscopic time scale below which the inertia of the heat bath becomes relevant. This can be represented by a high frequency cutoff in the memory friction function $\tilde{\gamma}(\omega)$, which in turn means that the cutoff being introduced in the spectral density of bath oscillators. Hence we regularize the momentum dispersion with the help of the Drude cutoff.
\subsubsection{Equilibrium momentum dispersion}
In order to calculate the equilibrium dispersion of the kinematic momentum of the dissipative charged oscillator in a magnetic field, we need the correlation function $\langle v_x ^2 + v_y ^2 \rangle$, and using which we write
\begin{equation}
\bigg\langle \bigg({\bf{p}}-\frac{e{\bf{A}}}{c}\bigg)^2 \bigg\rangle = m^2 \langle v_x ^2 + v_y ^2 \rangle - m\hbar\omega_c .
\end{equation}
After doing a contour integration in Eq.(\ref{velocity-correlation-function}), we obtain\cite{jishad2}
\begin{equation}
\begin{split}
\langle v_x ^2 + v_y ^2 \rangle &= \frac{2}{m\beta} + \frac{\hbar\omega_0 ^2}{m\pi} \sum_{j=1}^3 \bigg[ q_j \psi\bigg(1+\frac{\lambda_j}{\nu} \bigg) + q_j ^{\prime} \psi\bigg(1+\frac{\lambda_j ^{\prime}}{\nu} \bigg) \bigg] \\
&+ \frac{\hbar}{m \pi}\sum_{j=1}^3 \bigg[ p_j \psi \bigg(1+ \frac{\lambda_j}{\nu} \bigg)+ p^{\prime} _j \psi \bigg(1+ \frac{\lambda^{\prime} _j}{\nu} \bigg) \bigg] + \frac{\hbar\omega_c}{m} ,
\end{split}
\label{equilibrium-momentum-dispersion-first}
\end{equation}
where $q_j$s are defined in Eq.(\ref{first-check}), and 
\begin{equation}
p_1 = \frac{\lambda_1 [\gamma\omega_D - i\omega_c (\lambda_1 - \omega_D)]}{(\lambda_1 -\lambda_2)(\lambda_1 -\lambda_3)},
\end{equation}
$p_2$ and $p_3$ are obtained by a cyclic permutation of roots. Similarly $p_j ^{\prime}$s are obtained by priming the $\lambda_j$s in the above equation. We can rewrite the above equation in terms of $\langle {\bf{r}}^2 \rangle$ and is given by
\begin{equation}
\bigg\langle \bigg({\bf{p}}-\frac{e{\bf{A}}}{c}\bigg)^2 \bigg\rangle= m^2 \omega_0 ^2 \langle {\bf{r}}^2 \rangle + \Pi^2 ,
\label{momentum-position}
\end{equation}
where
\begin{equation}
\Pi^2 = \frac{m \hbar}{\pi}\sum_{j=1}^3 \bigg[p_j \psi \bigg(1+ \frac{\lambda_j}{\nu} \bigg)+ p^{\prime} _j \psi \bigg(1+ \frac{\lambda^{\prime} _j}{\nu} \bigg) \bigg].
\label{moment1}
\end{equation}

Expressing the digamma functions in a summation form, we write Eq.(\ref{momentum-position}) in terms of the bosonic Matsubara frequencies as
\begin{equation}
\bigg\langle \bigg({\bf{p}}-\frac{e{\bf{A}}}{c}\bigg)^2 \bigg\rangle = \frac{2m}{\beta} + \frac{4m}{\beta}  \sum_{n=1}^\infty \frac{(\nu_n ^2 + \omega_0 ^2 + \frac{\nu_n \gamma\omega_D}{(\nu_n +\omega_D)})(\omega_0 ^2 + \frac{\nu_n \gamma\omega_D}{(\nu_n +\omega_D)})+\omega_c ^2 \nu_n ^2}{(\nu_n ^2 + \omega_0 ^2 + \frac{\nu_n \gamma\omega_D}{(\nu_n +\omega_D)})^2 + \omega_c ^2 \nu_n ^2}.
\label{digamma-matsubara}
\end{equation}
At very low temperatures,
\begin{equation}
\Pi^2 = -\frac{2m\hbar \gamma \pi}{3}\bigg(\frac{1}{\beta\hbar\omega_0}\bigg)^2 + \frac{m\hbar}{\pi}\sum_{j=1}^3 [p_j \ln \lambda_j + p^{\prime} _j \ln \lambda^{\prime} _j],
\end{equation}
hence
\begin{equation}
\bigg\langle \bigg({\bf{p}}-\frac{e{\bf{A}}}{c}\bigg)^2 \bigg\rangle _{{\rm Low-T}} = m^2 \omega_0 ^2 \langle {\bf{r}}^2 \rangle _{{\rm Low-T}} + \Pi^2 _{{\rm Low-T}} =  \frac{m\hbar}{\pi}\sum_{j=1}^3 \bigg[ (\omega_0 ^2 q_j + p_j)\ln \lambda_j + (\omega_0 ^2 q^{\prime} _j + p^{\prime} _j)\ln \lambda^{\prime} _j \bigg] + \mathcal{O}(T^4),
\end{equation}
so that
\begin{equation}
\begin{split}
\bigg\langle \bigg({\bf{p}}-\frac{e{\bf{A}}}{c}\bigg)^2 \bigg\rangle _{T=0} &= m^2 \omega_0 ^2 \langle {\bf{r}}^2 \rangle _{T=0} + \frac{m\hbar}{\pi}\sum_{j=1}^3 \bigg[p_j \ln \lambda_j + p^{\prime} _j \ln \lambda^{\prime} _j \bigg]  \\
&= \frac{m\hbar}{\pi}\sum_{j=1}^3 \bigg[ (\omega_0 ^2 q_j + p_j)\ln \lambda_j   + (\omega_0 ^2 q^{\prime} _j + p^{\prime} _j)\ln \lambda^{\prime} _j \bigg] .
\end{split}
\end{equation}
We have already seen that the equilibrium position and momentum dispersions are expressed in terms of the imaginary part of the dynamical susceptibility $\tilde{\chi}^{\prime\prime} (\tilde{\omega})$. Because of the fact that $\tilde{\chi}^{\prime\prime} (\tilde{\omega})$ is an odd function in $\tilde{\omega}$, the asymptotic low-temperature expansions of position and momentum dispersions are power series expansions in $T^2$. For the Drude bath, leading term in the position dispersion is $T^2$, while the $T^2$ term in the momentum dispersion cancels out with a corresponding contribution from the term $m^2 \omega_0 ^2 \langle {\bf{r}}^2 \rangle $. Hence the leading contribution in the momentum dispersion varies as $T^4$.

In the limit of $\omega_D \rightarrow \infty$, at low-temperatures, $\Pi^2$ can be written as
\begin{equation}
\Pi^2 = \frac{2m\hbar\gamma}{\pi}\ln \bigg(\frac{\omega_D}{\omega_0} \bigg)-\frac{2\pi\gamma m\hbar}{3}\bigg(\frac{1}{\beta\hbar\omega_0} \bigg)^2 - \frac{m\pi}{\hbar}\bigg[\frac{\bar{\gamma}^2}{2\Lambda}\ln\bigg(\frac{\lambda_1}{\lambda_2}\bigg)+ \frac{\bar{\gamma}^{*2}}{2\Lambda^{\prime}}\ln\bigg(\frac{\lambda^{\prime} _1}{\lambda^{\prime} _2}\bigg)\bigg] + \mathcal{O}(T^4),
\end{equation}
where we have neglected the terms of the order of $\gamma / \omega_D$ and higher orders of $\omega_D ^{-1}$. This immediately results in
\begin{eqnarray}
\bigg\langle \bigg({\bf{p}}-\frac{e{\bf{A}}}{c}\bigg)^2 \bigg\rangle &=& m^2 \omega_0 ^2 \langle {\bf{r}}^2 \rangle + \Pi^2 \nonumber\\
&=& \frac{m\omega_0 ^2 \hbar}{\pi}\bigg[\frac{1}{\Lambda} \ln \bigg(\frac{\lambda_1}{\lambda_2}\bigg) + \frac{1}{\Lambda^{\prime}} \ln \bigg(\frac{\lambda^{\prime} _1}{\lambda^{\prime} _2}\bigg) \bigg] - \frac{m\hbar}{\pi}\bigg[\frac{\bar{\gamma}^2}{2\Lambda}\ln\bigg(\frac{\lambda_1}{\lambda_2}\bigg)+ \frac{\bar{\gamma}^{*2}}{2\Lambda^{\prime}}\ln\bigg(\frac{\lambda^{\prime} _1}{\lambda^{\prime} _2}\bigg)\bigg]\nonumber\\
&+&  \frac{2m\gamma \hbar}{\pi} \ln \bigg( \frac{\omega_D}{\omega_0} \bigg) + \mathcal{O}(T^4),
\label{momentum-way1}
\end{eqnarray}
so that
\begin{equation}
\bigg\langle \bigg({\bf{p}}-\frac{e{\bf{A}}}{c}\bigg)^2 \bigg\rangle _{T=0} = \frac{2m\hbar\gamma}{\pi}\ln \bigg(\frac{\omega_D}{\omega_0} \bigg) + \frac{m\hbar}{\pi}\bigg[\frac{1}{\Lambda} \bigg( \omega_0^2-\frac{\bar{\gamma}^2}{2}\bigg)\ln\bigg(\frac{\lambda_1}{\lambda_2}\bigg) + \frac{1}{\Lambda^{\prime}} \bigg(\omega_0 ^2-\frac{\bar{\gamma}^{*2}}{2}\bigg)\ln\bigg(\frac{\lambda^{\prime} _1}{\lambda^{\prime} _2}\bigg)\bigg] .
\end{equation}
Alternatively we write
\begin{equation}
\bigg\langle \bigg({\bf{p}}-\frac{e{\bf{A}}}{c}\bigg)^2 \bigg\rangle _{T=0} = m^2 \omega_0 ^2 \bigg(1-\frac{\gamma^2}{2\omega_0 ^2}+\frac{\omega_c ^2}{2\omega_0 ^2}\bigg)\langle {\bf{r}}^2 \rangle _{T=0} -\frac{i\omega_c m\hbar\gamma}{\pi}\bigg[\frac{1}{\Lambda}\ln\bigg(\frac{\lambda_1}{\lambda_2}\bigg) - \frac{1}{\Lambda^{\prime}}\ln\bigg(\frac{\lambda^{\prime} _1}{\lambda^{\prime} _2}\bigg)\bigg]+\frac{2m\hbar\gamma}{\pi}\ln \bigg(\frac{\omega_D}{\omega_0} \bigg) .
\end{equation}
We may say that the reservoir is continuously measuring the position of the charged particle in the oscillator, the oscillator is getting more localized in space but the kinematic momentum spread and as a result the kinetic energy of the system becomes larger as the damping is increased. For large damping $\bar{\gamma}\gg 2\omega_0$, we have from Eq.(\ref{momentum-way1})
\begin{eqnarray}
\bigg\langle \bigg({\bf{p}}-\frac{e{\bf{A}}}{c}\bigg)^2 \bigg\rangle &=& m^2 \omega_0 ^2 \langle {\bf{r}}^2 \rangle + \Pi^2 \nonumber\\
&=& \frac{2m\gamma \hbar}{\pi} \ln \bigg( \frac{\omega_D}{\omega_0} \bigg) + \frac{2m\omega_0 ^2 \hbar}{\pi}\bigg[\frac{1}{\bar{\gamma}} \ln \bigg(\frac{\bar{\gamma}}{\omega_0}\bigg) + \frac{1}{\bar{\gamma^*}} \ln \bigg(\frac{\bar{\gamma}^*}{\omega_0}\bigg) \bigg] - \frac{m\hbar}{\pi}\bigg[\bar{\gamma} \ln\bigg(\frac{\bar{\gamma}}{\omega_0}\bigg)+ \bar{\gamma}^{*} \ln\bigg(\frac{\bar{\gamma}^*}{\omega_0}\bigg)\bigg]\nonumber\\
&+& \mathcal{O}(T^4),
\end{eqnarray}
which in the absence of magnetic field ($\omega_c =0$) yields the result
\begin{equation}
\bigg\langle \bigg({\bf{p}}-\frac{e{\bf{A}}}{c}\bigg)^2 \bigg\rangle  = \frac{2m\hbar\gamma}{\pi}\ln \bigg(\frac{\omega_D}{\omega_0} \bigg) + \frac{4m\omega_0 ^2 \hbar}{\pi \gamma} \ln \bigg(\frac{\gamma}{\omega_0}\bigg) - \frac{2m\hbar\gamma}{\pi} \ln \bigg(\frac{\gamma}{\omega_0} \bigg) + \mathcal{O}(T^4)
\end{equation}
For very large damping $\gamma \gg \omega_0$ and $\gamma \gg \omega_c$ we have $\langle {\bf{r}}^2 \rangle _{T=0} = 4\hbar / m\pi\gamma \ln (\gamma / \omega_0)$ and $\bigg\langle \bigg({\bf{p}}-e{\bf{A}}/c \bigg)^2 \bigg\rangle _{T=0} = 2m\gamma\hbar\gamma /\pi \ln (\omega_D / \gamma)$. When $\hbar\omega_D \gg k_B T$, the leading contribution to $\Pi^2$ is
\begin{equation}
\Pi^2 = \frac{2m\hbar\gamma}{\pi}\ln \bigg(\frac{\beta\hbar\omega_D}{2\pi} \bigg).
\label{leading-correction}
\end{equation}
\begin{figure}
  \centering
  \subfloat[]{\label{fig:low-c-q}\includegraphics[width=0.45\textwidth]{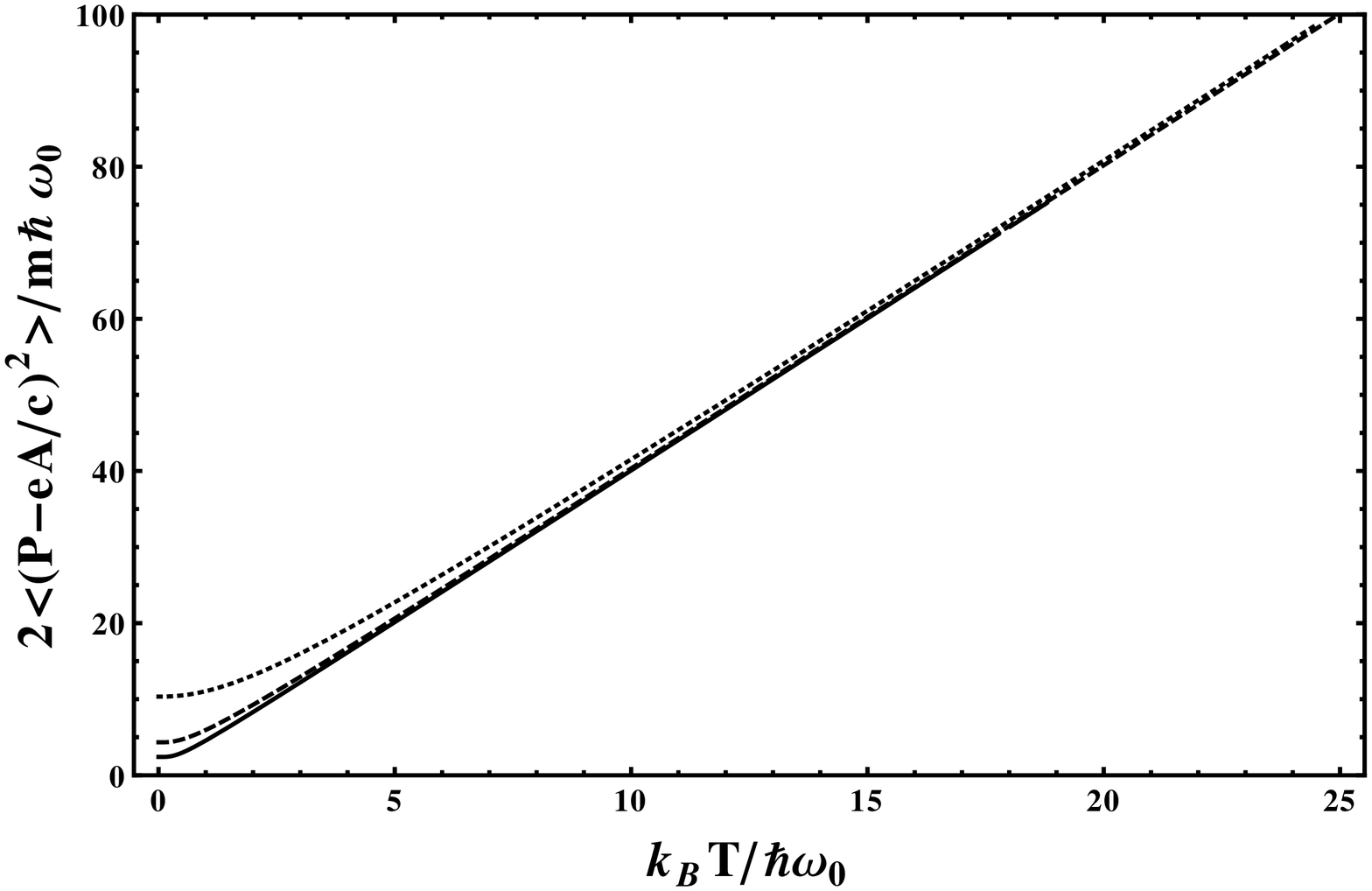}} \hspace {0.2cm}               
  \subfloat[]{\label{fig:high-c-q}\includegraphics[width=0.45\textwidth]{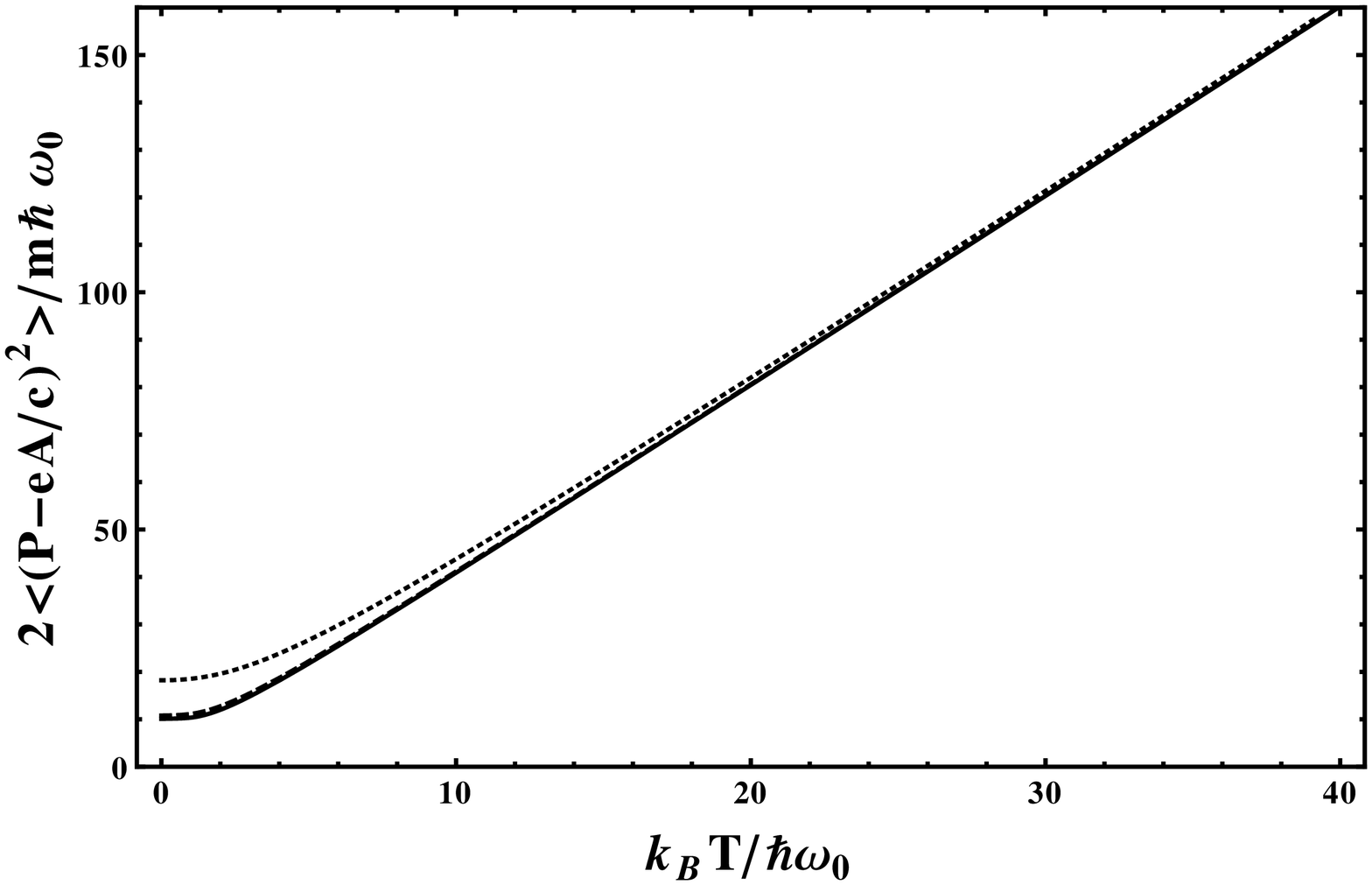}}
  \caption{The normalized equilibrium dispersion of momentum with respect to the dimensionless temperature $k_B T / \hbar\omega_0$ for (a) low magnetic field ($\omega_c / \omega_0 = 0.5$) and (b) for high magnetic field ($\omega_c / \omega_0 = 5.0$) is given above. The thick curve is for $\gamma / \omega_0 = 0.1$, the dashed one is for $\gamma / \omega_0 = 1.0$ and dotted one is for $\gamma / \omega_0 = 5.0$. Here we set $\omega_D / \omega_0 = 10.0$.}
  \label{avg-moment}
\end{figure}
At extremely high temperatures, i.e., for $k_B T$ greater than any other scales in the system, upon expanding the digamma functions, we obtain
\begin{equation}
\Pi^2 = \frac{m\beta\hbar^2}{6}(\gamma\omega_D + \omega_c ^2),
\end{equation}
$\Pi^2$ is negligibly small at high temperatures. Hence the correct classical result is recovered for $\langle ({\bf{p}}-\frac{e{\bf{A}}}{c})^2 \rangle$, i.e., $\langle ({\bf{p}}-\frac{e{\bf{A}}}{c})^2 \rangle = m^2 \omega_0 ^2 \langle {\bf{r}}^2 \rangle = \frac{2m}{\beta}$. 

The temperature dependence of the position dispersion is given in figure 1 and that of the momentum dispersion is shown in figure 2, for both low and high magnetic field strengths respectively. The results are as expected from the theory. At low temperatures, as the damping increases, the position dispersion decreases while the momentum dispersion increases. But at high temperatures, both the dispersions become independent of the dissipation constant $\gamma$ and vary with $T$ linearly, thereby proving the classical equipartition result. At low temperatures and at a higher magnetic field strength, the dispersion in the position remains at a small value even for higher dissipation strengths. This proves the fact that strong dissipation localizes the motion and the magnetic field enhances the effect. Therefore we may call the term $\gamma (t)/m+i\omega_c$ as an \textit{effective damping} term.
\subsubsection{Equilibrium momentum dispersion of the free particle}
In this subsection we derive the equilibrium momentum dispersion of a free damped charged particle in the presence of a magnetic field. We can start with Eq.(\ref{velocity-correlation-function}) and apply the limit $\omega_0 \rightarrow 0$ to calculate the velocity autocorrelation function for the free (unconfined) system. Alternatively one can obtain the free particle momentum dispersion from Eq.(\ref{equilibrium-momentum-dispersion-first}) as well in the limit of $\omega_0 \rightarrow 0$. It therefore gives
\begin{equation}
\bigg\langle \bigg({\bf{p}}-\frac{e{\bf{A}}}{c}\bigg)^2 \bigg\rangle = \frac{2m}{\beta}+\frac{m\hbar}{\pi}\bigg\{p_2 \psi\bigg(1+\frac{\lambda_2}{\nu} \bigg)+p_3 \psi\bigg(1+\frac{\lambda_3}{\nu} \bigg) + p^{\prime} _2 \psi\bigg(1+\frac{\lambda^{\prime} _2}{\nu} \bigg)+p^{\prime} _3 \psi\bigg(1+\frac{\lambda^{\prime} _3}{\nu} \bigg) \bigg\},
\label{momentum-unconfined}
\end{equation}
where
\begin{equation}
\begin{split}
p_2 &= \frac{[\gamma\omega_D - i\omega_c (\lambda_2 - \omega_D)]}{(\lambda_2  - \lambda_3)}, \\
p_3 &= -\frac{[\gamma\omega_D - i\omega_c (\lambda_3 - \omega_D)]}{(\lambda_2  - \lambda_3)},
\end{split}
\end{equation}
and the corresponding primed values are obtained by replacing $-i\omega_c$ with its complex conjugate and $\lambda_j$s with their corresponding primed ones. As we discussed earlier, in the limit of $\omega_0 \rightarrow 0$, one of the roots vanish from the vieta equations so that we have $\lambda_2 + \lambda_3  = \omega_D + i\omega_c,~~\lambda_2 \lambda_3  = \omega_D (\gamma + i\omega_c)$ and $\lambda^{\prime} _2 + \lambda^{\prime} _3  = \omega_D - i\omega_c,~~\lambda^{\prime} _2 \lambda^{\prime} _3  = \omega_D (\gamma - i\omega_c)$. In the limit of $\omega_D \rightarrow \infty$, we have $\lambda_2 \sim \omega_D,~~\lambda_3 \sim \gamma + i\omega_c$ and $\lambda^{\prime} _2 \sim \omega_D,~~\lambda^{\prime} _3 \sim \gamma - i\omega_c$. Hence in the limit of $\omega_D \rightarrow \infty$ we obtain
\begin{equation}
\bigg\langle \bigg({\bf{p}}-\frac{e{\bf{A}}}{c}\bigg)^2 \bigg\rangle = \frac{2m}{\beta}+\frac{m\hbar}{\pi}\bigg\{2\gamma \ln \bigg(\frac{\omega_D}{\nu}\bigg) - \bar{\gamma}\psi\bigg(1+\frac{\bar{\gamma}}{\nu} \bigg) -\bar{\gamma}^* \psi\bigg(1+\frac{\bar{\gamma}^*}{\nu}\bigg) \bigg\},
\label{moment-dissipation-free}
\end{equation}
where $\bar{\gamma} = \gamma + i\omega_c$ and $\bar{\gamma}^*$ its complex conjugate. Here $\nu$ is the first bosonic Matusbara frequency defined earlier. We can alternatively obtain the momentum dispersion of the free damped charged particle in a magnetic field from Eq.(\ref{digamma-matsubara}) in a summation form by taking the limit $\omega_0 \rightarrow 0$ and is given by
\begin{equation}
\bigg\langle \bigg({\bf{p}}-\frac{e{\bf{A}}}{c}\bigg)^2 \bigg\rangle = \frac{2m}{\beta} + \frac{4m}{\beta}  \sum_{n=1}^\infty \frac{(\nu_n ^2 + \frac{\nu_n \gamma\omega_D}{(\nu_n +\omega_D)})\frac{\nu_n \gamma\omega_D}{(\nu_n +\omega_D)}+\omega_c ^2 \nu_n ^2}{(\nu_n ^2 + \frac{\nu_n \gamma\omega_D}{(\nu_n +\omega_D)})^2 + \omega_c ^2 \nu_n ^2}.
\label{digamma1-matsubara}
\end{equation}
Just as in the case of a damped charged oscillator in a magnetic field, for strict ohmic damping, the equilibrium momentum dispersion for the damped free particle in a magnetic field also diverges. This divergence is eliminated by regularizing it with the Drude cutoff in the spectral density of the bath modes.

In the zero damping case ($\gamma = 0$), from Eq.(\ref{moment-dissipation-free}), after simplifications, we write
\begin{equation}
\bigg\langle \bigg({\bf{p}}-\frac{e{\bf{A}}}{c}\bigg)^2 \bigg\rangle = m\hbar\omega_c \coth\bigg(\frac{\pi \omega_c}{\nu}\bigg) .
\label{moment-free-field}
\end{equation}
In obtaining this result we have used the formula
\begin{equation}
\psi\bigg(\frac{i\omega_c}{\nu} \bigg) - \psi\bigg(\frac{-i\omega_c}{\nu} \bigg) = -\frac{\pi}{i}\coth\bigg(\frac{\pi\omega_c}{\nu} \bigg) - \frac{\nu}{i\omega_c}.
\end{equation}
Using Eq.(\ref{moment-free-field}), we obtain the energy
\begin{equation}
E = \frac{1}{2m}\bigg\langle \bigg({\bf{p}}-\frac{e{\bf{A}}}{c}\bigg)^2 \bigg\rangle = \frac{\hbar\omega_c}{2}\coth\bigg(\frac{\pi \omega_c}{\nu}\bigg),
\end{equation}
for a free charged particle in a magnetic field in the absence of dissipation. Meanwhile in the absence of the magnetic field ($\omega_c  =0$), from Eq.(\ref{moment-dissipation-free}) we get
\begin{equation}
\langle {\bf{p}}^2 \rangle = \frac{2m}{\beta}+\frac{2 m\hbar \gamma}{\pi}\bigg\{\ln \bigg(\frac{\omega_D}{\nu}\bigg) - \psi\bigg(1+\frac{\gamma}{\nu} \bigg) \bigg\}.
\label{check25}
\end{equation}
On the other hand, for Drude bath, when $\omega_c  =0$, we have the roots $\lambda_{2,3} = \omega_D /2 \pm \sqrt{\omega_D ^2 /4 -\gamma\omega_D}$ and using which we write, from Eq.(\ref{momentum-unconfined}),
\begin{equation}
\langle {\bf{p}}^2 \rangle  = \frac{2m}{\beta} + \frac{2m\hbar\gamma}{\pi} \frac{\omega_D}{\sqrt{\omega_D ^2 - 4\gamma\omega_D}}\bigg\{\psi\bigg(1+\frac{\lambda_2}{\nu} \bigg) +\psi\bigg(1+\frac{\lambda_3}{\nu} \bigg) \bigg\} ,
\label{free-particle-momentum-result}
\end{equation}
which is exactly matching with the free particle result in two dimensions. At high temperatures, the first term in Eq.(\ref{free-particle-momentum-result}) dominates. Note here that in the high frequency cutoff limit ($\omega_D \rightarrow \infty$), Eq.(\ref{free-particle-momentum-result}) reduces to Eq.(\ref{check25}).

The low temperature behavior of Eq.(\ref{moment-dissipation-free}) can be easily obtained using the asymptotic expansion of the digamma functions 
\begin{equation}
\psi(z)=\ln z -\frac{1}{2z}-\frac{1}{12 z^2}+\frac{1}{120 z^4} ,
\end{equation}
and it turns out to be 
\begin{equation}
\bigg\langle \bigg({\bf{p}}-\frac{e{\bf{A}}}{c}\bigg)^2 \bigg\rangle = \frac{2\pi\gamma m\hbar}{3}\bigg( \frac{1}{\beta\hbar\sqrt{\gamma^2 + \omega_c ^2}}\bigg)^2 + \frac{m\hbar\gamma}{\pi}\ln\bigg(\frac{\omega_D ^2}{\gamma^2 + \omega_c ^2}\bigg)-\frac{im\hbar\omega_c}{\pi}\ln\bigg(\frac{\gamma + i\omega_c}{\gamma - i\omega_c}\bigg)
\end{equation}
At zero temperature ($T=0$) and for $\omega_c  =0$, we have 
\begin{equation}
\langle {\bf{p}}^2 \rangle  = \frac{2m\hbar\gamma}{\pi} \ln \bigg(\frac{\omega_D}{\gamma}\bigg).
\end{equation}
This result can be obtained also from Eq.(\ref{free-particle-momentum-result}) at $T=0$ with the condition $\omega_D \gg 4\gamma$. Meanwhile at high temperatures, the mean squared kinematic momentum of the damped charged particle in a magnetic field behaves as
\begin{equation}
\bigg\langle \bigg({\bf{p}}-\frac{e{\bf{A}}}{c}\bigg)^2 \bigg\rangle = \frac{2m}{\beta}+\frac{2m\hbar\gamma}{\pi}\ln\bigg(\frac{\omega_D}{\nu}\bigg)+\frac{2m\hbar\gamma}{\pi}\gamma_E - \frac{m\beta\hbar^2}{6}(\gamma^2 - \omega_c ^2),
\end{equation} 
where $\gamma_E$ is the Euler gamma which is a constant with a value $\sim 0.577$. To obtain the above result, we have used the expansion of the digamma function $\psi(z)=-\frac{1}{z}-\gamma_E + \frac{\pi^2}{6}z$. At high enough temperatures, we see that the equipartition theorem is satisfied and the leading correction term to the kinematic momentum dispersion is exactly similar to that in Eq.(\ref{leading-correction}), for the confined system.
\subsubsection{Partition function and the dispersions}
It is also possible to obtain the equilibrium position and momentum dispersions from the reduced partition function of the system. The reduced partition function is defined in terms of the partition functions of the coupled system and the uncoupled bath \cite{ingoldbook,leggett,schramm,talkner,ingold-acta,ford,feynman,feynman1,garg,lewis,hanke,ingold-lect}, which is defined as
\begin{equation}
\mathcal{Z}_R (\beta) = \frac{{\rm Tr_{S+B}}[\exp(-\beta\mathcal{H})]}{{\rm Tr_B} [\exp(-\beta\mathcal{H}_B)]} = \frac{\mathcal{Z} ^{\rm Total}}{\mathcal{Z}_B},
\end{equation}
where $\mathcal{H}$ is the total Hamiltonian consisting of contributions from the system, bath and the interaction, $\mathcal{Z} ^{\rm Total}$ represents the partition function of the composite system and $\mathcal{Z}_B$ represents the partition function of the heat bath. For a dissipative charged harmonic oscillator in a uniform and homogeneous magnetic field, the relations connecting the mean squared values of the position and the kinematic momentum and the partition function are
\begin{eqnarray}
\langle {\bf{r}}^2 \rangle &=& -\frac{1}{m\beta \omega_0}\frac{d}{d\omega_0}\ln \mathcal{Z}_R (\beta),\\
\bigg\langle \bigg({\bf{p}}-\frac{e{\bf{A}}}{c}\bigg)^2 \bigg\rangle &=& -\frac{m}{\beta}\bigg[ \omega_0 \frac{d}{d\omega_0} + 2\gamma \frac{d}{d\gamma} + 2\omega_c \frac{d}{d\omega_c} \bigg]\ln \mathcal{Z}_R (\beta),\label{momentum-dispersion-partition-function}
\end{eqnarray} 
where the reduced partition function $\mathcal{Z}_R (\beta)$ can be calculated using an imaginary path integral technique. In Eq.(\ref{momentum-dispersion-partition-function}), we have employed the representation $\hat{\gamma}(z)=\gamma g(z)$. Note here that the expressions given above are in general valid for any form of linear memory friction. The partition function for the charged oscillator in a magnetic field in the presence of a finite dissipative quantum heat bath can be written as \cite{jishad1} 
\begin{equation}
\mathcal{Z}_R (\beta) = \bigg(\frac{1}{\beta\hbar\omega_0} \bigg)^2 \prod_{n=1}^\infty \frac{\nu_n ^4}{(\nu_n ^2 +\omega_0 ^2 + \nu_n \hat{\gamma}(\nu_n))^2 + \omega_c ^2 \nu_n ^2} ,
\label{partition-function-matsubara}
\end{equation}
which, with the help of the Drude cut off and the definition of the Gamma function, can be expressed as
\begin{equation}
\mathcal{Z}_R (\beta) = \bigg(\frac{\beta\hbar\omega_0}{4\pi^2} \bigg)^2 \frac{\prod_{j=1}^3 \Gamma\bigg(\frac{\beta\hbar\lambda_j}{2\pi} \bigg) \Gamma\bigg(\frac{\beta\hbar\lambda_j ^{\prime}}{2\pi} \bigg)}{\bigg(\Gamma\bigg(\frac{\beta\hbar\omega_D}{2\pi} \bigg) \bigg)^2} ,
\label{partition-function-gamma}
\end{equation}
where the $\lambda_j s$ and $\lambda_j ^{\prime} s$ satisfy the same vieta equations given in Eq.(\ref{vieta}) and Eq.(\ref{vieta1}).

We can immediately calculate the average (mean) energy $E$ at equilibrium from the equilibrium dispersions of the position and the kinematic momentum which results in
\begin{equation}
E = \langle \mathcal{H}_S \rangle = \frac{1}{2m}\bigg\langle \bigg({\bf{p}}-\frac{e{\bf{A}}}{c}\bigg)^2 \bigg\rangle +\frac{1}{2}m\omega_0 ^2 \langle {\bf{r}}^2 \rangle ,
\end{equation}
and it turns out to be the quantity
\begin{equation}
E = \frac{2}{\beta}+\frac{\hbar}{2\pi}\sum_{j=1}^3 \bigg\{[2\omega_0 ^2 q_j + p_j ]\psi\bigg(1+\frac{\lambda_j}{\nu} \bigg) + [2\omega_0 ^2 q_j ^{\prime} + p_j ^{\prime} ]\psi\bigg(1+\frac{\lambda_j ^{\prime}}{\nu} \bigg) \bigg\} .
\end{equation}
Note that the mean energy is being calculated using the well known formula in statistical physics\cite{ingold-acta,talkner} 
\begin{equation}
E = \langle \mathcal{H}_S \rangle = \frac{{\rm Tr_{S+B}}[\mathcal{H}_S \exp(-\beta\mathcal{H})]}{{\rm Tr_{S+B}} [\exp(-\beta\mathcal{H})]}.
\end{equation}
In the strict ohmic limit of $\omega_D \rightarrow \infty$, one of the roots, say, $\lambda_3 = \lambda_3 ^{\prime} \sim \omega_D$ and other roots are $\lambda_{1,2} = \bar{\gamma}/2 \pm 1/2 \sqrt{\bar{\gamma}^2 - 4\omega_0 ^2}$. Also $\lambda_{1,2} ^{\prime} = \bar{\gamma}^* /2 \pm 1/2 \sqrt{\bar{\gamma}^{*2} - 4\omega_0 ^2}$. Moreover, $q_1 = 1/(\lambda_1 -\lambda_2) = -q_2$, $q_3 =0$ and $p_1 = -[\lambda_1 (\bar{\gamma})]/(\lambda_1 -\lambda_2)$, $p_2  = [\lambda_2 (\bar{\gamma})]/(\lambda_1 -\lambda_2)$, $p_3 \sim \gamma$. For $q_j ^{\prime}$'s and $p_j ^{\prime}$'s we need to replace $\bar{\gamma}$ with $\bar{\gamma}^*$ and $\lambda_j$'s with $\lambda^{\prime} _j$'s in the corresponding values of $q_j$'s and $p_j$'s. Using these substitutions and with the recurrence formula $\psi(1+z) = \psi(z) +1/z$ of the digamma function, we write the average energy as
\begin{equation}
E = -\frac{2}{\beta} -\frac{1}{\beta} \bigg\{\frac{\lambda_1}{\nu}\psi\bigg(\frac{\lambda_1}{\nu} \bigg) + \frac{\lambda_2}{\nu}\psi\bigg(\frac{\lambda_2}{\nu} \bigg) + \frac{\lambda_1 ^{\prime}}{\nu}\psi\bigg(\frac{\lambda_1 ^{\prime}}{\nu} \bigg)+ \frac{\lambda_2 ^{\prime}}{\nu}\psi\bigg(\frac{\lambda_2 ^{\prime}}{\nu} \bigg) \bigg\} ,
\label{average-energy-langevin}
\end{equation}
which in the absence of dissipation ($\gamma =0$) gives
\begin{equation}
E = \bigg(\frac{\hbar\omega_+}{2} \bigg)\coth \bigg(\frac{\beta\hbar\omega_+}{2} \bigg) + \bigg(\frac{\hbar\omega_-}{2} \bigg)\coth \bigg(\frac{\beta\hbar\omega_-}{2} \bigg) ,
\end{equation}
which is nothing but the internal energy of the charged harmonic oscillator in a magnetic field (or the Fock-Darwin model) and we define $\omega_{\pm} = \sqrt{\omega_0 ^2 + \omega_c ^2 /4} \pm \omega_c /2$. Now from the partition function in Eq.(\ref{partition-function-gamma}) we can calculate the internal energy $U$ using the relation 
\begin{equation}
U=-\frac{\partial}{\partial \beta} \ln \mathcal{Z}_R (\beta) ,
\end{equation}
and is given by
\begin{equation}
U = -\frac{2}{\beta}-\frac{1}{\beta}\sum_{j=1}^3 \bigg\{\frac{\lambda_j}{\nu}\psi\bigg(\frac{\lambda_j}{\nu} \bigg) +\frac{\lambda_j ^{\prime}}{\nu}\psi\bigg(\frac{\lambda_j ^{\prime}}{\nu} \bigg) \bigg\} + \frac{2}{\beta}\frac{\omega_D}{\nu}\psi\bigg(\frac{\omega_D}{\nu} \bigg).
\end{equation}
In the strict ohmic limit we see that the internal energy $U$ is exactly same as that in Eq.(\ref{average-energy-langevin}) for $E$. The root cause of this intuitive similarity between the two approaches can be easily understood by the following simple derivations. Using the Matsubara representation of the position and the kinematic momentum, we can write the average energy in a closed form given by
\begin{equation}
E = \frac{2}{\beta} + \frac{2}{\beta} \sum_{n=1}^{\infty} \frac{(\nu_n ^2 +\omega_0 ^2 + \nu_n \hat{\gamma}(\nu_n))(2\omega_0 ^2 + \nu_n \hat{\gamma}(\nu_n)) + \omega_c ^2 \nu_n ^2}{(\nu_n ^2 +\omega_0 ^2 + \nu_n \hat{\gamma}(\nu_n))^2 + \omega_c ^2 \nu_n ^2} .
\label{energy-mean-summation}
\end{equation}
Similarly we can obtain the internal energy $U$ from the partition function in Eq.(\ref{partition-function-matsubara}) and is given by
\begin{equation}
U = \frac{2}{\beta} + \frac{2}{\beta} \sum_{n=1}^{\infty} \frac{(\nu_n ^2 +\omega_0 ^2 + \nu_n \hat{\gamma}(\nu_n))(2\omega_0 ^2 + \nu_n \hat{\gamma}(\nu_n)-\nu_n ^2 \hat{\gamma}^{\prime} (\nu_n)) + \omega_c ^2 \nu_n ^2}{(\nu_n ^2 +\omega_0 ^2 + \nu_n \hat{\gamma}(\nu_n))^2 + \omega_c ^2 \nu_n ^2} .
\label{internal-energy-mean-summation}
\end{equation}
It is clear from the Eqs.(\ref{energy-mean-summation}) and (\ref{internal-energy-mean-summation}) that they differ actually due to a term $-\nu_n ^2 \hat{\gamma}^{\prime} (\nu_n)$ in the numerator in the frequency dependent damping case. But for the strict ohmic case where $\hat{\gamma}(\nu_n) = \gamma$, the derivative is zero and that renders the two results same. This proves that the issue of this ``equality" seem to be just esoteric to the strict ohmic damping case. In our previous work \cite{jishad2}, we have seen that in the strict ohmic limit ($\omega_D \rightarrow \infty$), the specific heat calculated from the two different approaches matches each other exactly. This rather puzzling equality of the two approaches in the strict ohmic limit was pointed out earlier by H\"anggi and Ingold\cite{ingold-acta} and H\"anggi et al.,\cite{talkner}, for a damped harmonic oscillator and a damped quantum free particle respectively. We therefore emphasize here the same conclusion that there is no good reason why these two results obtained through two different approaches match exactly in the memoryless damping limit.
\section{Conclusion}
\label{sec:11}
We have studied in detail the position and the kinematic momentum autocorrelation functions of a damped charged harmonic oscillator in a magnetic field. The fluctuation-dissipation relation in the context of dissipative Landau diamagnetism have been verified. Moreover, we have elucidated the equilibrium dispersions in position and momentum and studied both the low and high temperature behaviors with and without the Drude cut-off frequency. At high enough temperatures, both the equilibrium dispersions are in accordance with the classical equipartition theorem. The numerical results for the dispersions are in accordance with the theoretical findings. We have verified the classical diffusive nature of the free charged particle in a magnetic field when the confinement frequency is turned off. Also in the limit of the vanishing confinement frequency $\omega_0$, we have obtained the equilibrium momentum dispersion of the free damped charged particle in a magnetic field. We also have shown the puzzling similarity in the energies when calculated from (i) the average of the effective stochastic Hamiltonian and from (ii) the thermodynamic partition function, under the often made Markovian assumption of memoryless damping.
\section{Acknowledgements}
Part of this work was completed when I was at the Saha Institute of Nuclear Physics (SINP), Kolkata, India and I am indebted to SINP for financial support during my stay. Part of this work was completed at the Institute of Physics (FUUK), Faculty of Mathematics and Physics, Charles University in Prague. I thank FUUK for the fantastic hospitality and financial support. I thank Prof. Dr. Gert.-Ludwig Ingold, Prof. Sushanta Dattagupta, and Dr. Subhasis Sinha for useful discussions and suggestions on many subtle points.

\end{document}